\documentclass[slac_one]{revtex4}
\usepackage{graphicx}
\usepackage{fancyhdr}
\usepackage{xspace}
\usepackage{multirow}
\def\babar{\mbox{\sl B\hspace{-0.37em} {\small\sl A}\hspace{-0.3em}
    \sl B\hspace{-0.37em} {\small\sl A\hspace{-0.02em}R}}}
\def\sbabar{\mbox{{\fontsize{9}{11}\sl B\hspace{-0.4em}
      \fontsize{7pt}{11pt}\sl A}\hspace{-0.35em} \fontsize{9}{11}\sl B\hspace{-0.4em} {\fontsize{7pt}{11pt}\sl
      A\hspace{-0.02em}R}}}
\def\CP {\ensuremath{C\!P}\xspace}
\def\mes        {\mbox{$m_{\rm ES}$}\xspace}

\begin{document}

\title{\boldmath Search for \CP violation and new physics in rare $B$ decays at the $B$ factories}

\author{G. Marchiori (\url{giovanni.marchiori@lpnhe.in2p3.fr}), 
  on behalf of the \babar\ and Belle Collaborations}
\affiliation{Laboratoire de Physique Nucl\'eaire et de Hautes Energies
  (IN2P3/CNRS), 4 Place Jussieu, 75005 Paris, France}

%
%Abstract
%
\begin{abstract}
We summarize recent results by the \sbabar\ and Belle Collaborations on the search for \CP
violation and new physics beyond the Standard Model in several rare $B$ meson decays.
The measurements exploit the final \sbabar\ and Belle $\Upsilon(4S)$ dataset, which consist
of approximately 470 and 772 million $B\overline{B}$ pairs, respectively.
\end{abstract}

\maketitle

\thispagestyle{fancy}


%
\section{INTRODUCTION} % Section title should be in all capitals.
The $B$ factory experiments, \babar\ at SLAC and Belle at KEK, have been operated 
successfully in the first decade of the 21st century, collecting in total more than one billion 
pairs of charged $(B^+B^-)$ and neutral $(B^0\overline{B}^0)$ $B$ mesons. 
Such large samples have been obtained from the decays of boosted $\Upsilon(4S)$ mesons 
produced in asymmetric $e^+e^-$ collisions at $\sqrt{s} = m_\Upsilon$, $e^+e^- \to \Upsilon(4S) \to B\overline{B}$.
After establishing experimentally $\CP$ violation in the $B$ meson system and confirming the 
Cabibbo-Kobayashi-Maskawa (CKM) paradigm~\cite{bib:Cabibbo,bib:KM} 
by measuring with good or excellent precision the sides and angles of the unitarity 
triangle in the ``golden" modes, like $B\to c\bar c K^{(*)0}$ for the angle $\beta = \arg[-V_{cd}V_{cb}^*/V_{td}V_{tb}^*]$ ($\sin 2\beta = 0.679 \pm 0.020$~\cite{bib:HFAG}), a significant fraction of the $B$ factories' physics 
program has been focused on rare $B$ decays, which can provide additional constraints on the 
unitarity triangle parameters, or can probe the presence of new physics (NP) 
beyond the Standard Model (SM).

\section{EXPERIMENTAL TECHNIQUES}
Both \babar\ and Belle are multi-purpose, (longitudinally) asymmetric detectors with nearly $4\pi$ 
coverage.
In the $e^+e^-$ collisions the $\Upsilon(4S)$ are produced with a Lorentz boost $\gamma\beta$
along the beam axis of 0.56 in \babar\ and 0.42 in Belle, resulting in an average longitudinal 
separation between the two $B$ decay vertices of $\approx$ 200-250 microns, from which the 
proper time difference between the two decays can be inferred.
The two detectors consist of an inner silicon detector to reconstruct $B$ and $D$ decay vertices
and low-$p_T$ charged particles, a gaseous drift chamber for charged particle tracking, various
devices optimized for excellent charged particle identification (PID) at low energy ($\approx 1$ GeV) 
through 
various techniques ($dE/dx$, Cherenkov light, time of flight, ..), homogeneous scintillating 
electromagnetic calorimeters for $e/\gamma$ reconstruction and identification, and muon and $K_L$ 
detectors installed in the iron yoke used to contain the flux return of the solenoid that provides the 
magnetic field for tracking. 

$B\overline{B}$ events produce typically a few number of charged ($\approx 11$) and neutral 
particles and often it is possible to fully reconstruct a $B$ decay, {\em i.e.} to measure directly
each of its decay products. This allows the calculation of two kinematic quantities
characterizing the $B$ meson,
the energy-substituted invariant mass (\mes or $M_{bc}$), 
$m_{\rm ES} = \sqrt{(E^*_{\rm beam})^2 - (p^*_B)^2}$, and 
the energy difference, $\Delta E = E^*_B - E^*_{\rm beam}$. For correctly reconstructed $B$ decays,
these two quantities peak around the $B$ meson mass ($m_B$) and zero, respectively.
Sometimes the $B$ decay under study produces ``invisible" particles that do not interact with the 
detector, for instance a neutrino or an hypothetical weakly interacting supersymmetric particle:
in that case, one can still fully reconstruct the recoiling $B$ (usually denoted as $B_{\rm tag}$)
and use it to infer the four-momentum of the signal $B$ and of the invisible particles.

When a pair of neutral $B$ mesons is produced, the flavors of the two mesons are entangled.
If the first $B$ meson decays at time $t$ to a flavor specific final state, which can be identified through 
the presence of leptons, kaons or soft pions, the second $B$ meson is inferred to have at the same 
time $t$ the opposite flavor. These correlations are exploited in time-dependent \CP violation measurements.

The $e^+e^-$ collisions can also produce light quark-antiquark pairs, $q\bar{q}$ ($q=u,d,s,c$).
The $B\overline{B}$ production cross section is around one fifth of the total hadronic cross section.
The hadronic $q\bar{q}$ backgrounds are typically distinguished from $B\overline{B}$ events by 
means of event-shape variables, that discriminate between the ``jet-like" topology of $q\bar{q}$ events
and the more ``spherical" $B\overline{B}$ events, and also exploit the different angular distributions
expected from angular momentum conservation and spin arguments. Because of the large 
correlations between the shape variables, those are usually combined in a single discriminating
quantity using multivariate techniques (Fisher discriminants, neural networks), trained on 
clean samples of $q\bar{q}$ events collected at $\sqrt{s}\approx 40-50$ MeV below the $\Upsilon(4S)$ peak
and with simulated signal samples.

\section{SEARCH FOR \CP VIOLATION IN CHARMLESS $B$ DECAYS}
Charmless hadronic $B$ decays provide sensitive probes for
\CP violation in the Standard Model and beyond, since:
\begin{itemize}
\item large \CP-violating effects are predicted by the SM in some of 
  these channels, and can therefore be measured with good precision;
\item the amount of \CP violation in some charmless $B$ decays is
  related, in the SM, to the amount of \CP violation -- measured with 
  excellent precision -- in $B^0 \to (c\bar c) K^0$ decays;
\item some of these decays proceeds mainly (or purely) through
  loop-mediated diagrams and SM predictions could therefore be
  significantly affected by possible new physics 
  contributions appearing in the loops of these diagrams.
\end{itemize}

\subsection{Search for \CP violation in two-body $B\to \eta h$ $(h=\pi^+,K^+,K^0)$ decays}
\begin{figure*}[t]
\centering
\includegraphics[width=135mm]{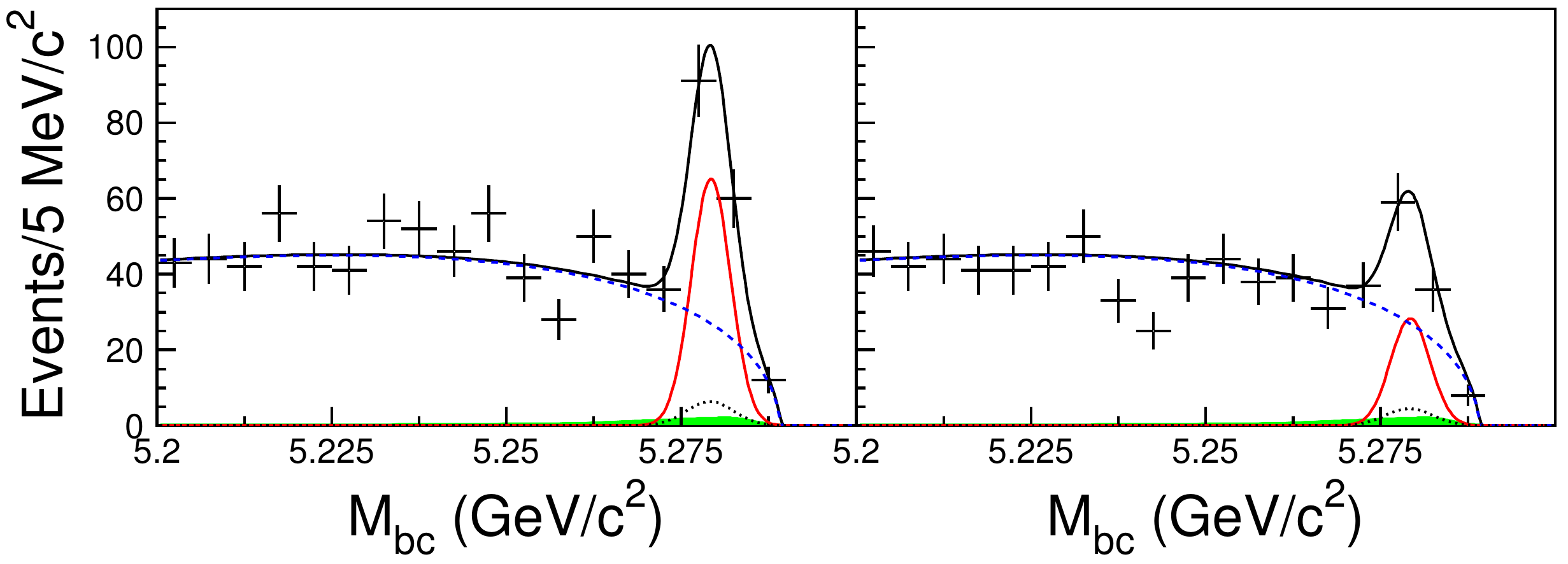}
\caption{$M_{bc}$ projections for $B^+ \to \eta K^+$ (left) and $B^- \to \eta K^-$ (right) candidate events selected by Belle~\cite{bib:Belle:etah}. Points with error bars represent the data, the total fit functions are shown by black solid curves, signals are shown by red solid curves, dashed curves are 
the continuum contributions, dotted curves are the cross-feed backgrounds from misidentification
and filled histograms are the contributions
from charmless $B$ backgrounds. The projections  are for events that have
$-0.10 <\Delta E <0.08$ GeV and $R^\prime > 1.95$.} \label{fig:Belle:etah}
\end{figure*}

Calculations of the SM contributions to the $B\to \eta K$ and $B\to \eta \pi$ decays, combined 
with the measured values of the $B^\pm\to \eta^{(\prime)}K^\pm$ and $B^0\to \eta^{\prime}K^0$
branching fractions, lead to the following expectations
\cite{bib:theo:etah1,bib:theo:etah2,bib:theo:etah3,bib:theo:etah4,bib:theo:etah5,bib:theo:etah6,bib:theo:etah7,bib:theo:etah8,bib:theo:etah9}:
\begin{enumerate}
\item the branching fraction of $B^0 \to \eta K^0$ is expected to
be lower than that of $B^\pm\to \eta K^\pm$, because the tree diagram in the 
$B^0 \to \eta K^0$ decay is color suppressed.
\item in $B^\pm \to \eta K^\pm$, a large direct \CP asymmetry $(A_{\CP})$ is expected,
as a consequence of the interference between the destructive combination of penguin amplitudes
and the tree amplitude.
\item in $B^\pm \to \eta \pi^\pm$, a large $A_{\CP}$ is also anticipated, due to the interference 
between the $b \to d$ penguin and $b\to u$ tree diagrams.
\end{enumerate}
Belle has published in~\cite{bib:Belle:etah} an updated measurement of 
$B\to \eta h$ $(h=\pi^+, K^+, K^0)$ decays\footnote{charge conjugate modes are implicitly 
included unless otherwise stated} using the final data sample of 
$772\times 10^6$ $B\overline{B}$ pairs. The $\eta$ meson is reconstructed in both the
$\pi^+\pi^-\pi^0$ and $\gamma\gamma$ final states, while the $K^0$ meson is selected
as a $K^0_S$ decaying to two oppositely-charged pions.
The full $B$ decay tree is reconstructed, with an efficiency -- including secondary branching 
fractions -- of around 1.5\% (4\%) for the $\eta K^0$ and 5\% (14\%) for the $\eta h^+$ final states, 
in the $\eta\to\pi^+\pi^-\pi^0$ ($\eta\to\gamma\gamma$) channel.
The signal yield is obtained through a maximum likelihood (ML) fit to $M_{bc}$, $\Delta E$ and 
an event-shape variable $R^\prime$. The samples containing the 
$\eta K^+$ and $\eta \pi^+$ candidates are fitted simultaneously, in order to constrain the
background from the $\eta K \leftrightarrow \eta \pi$ cross-feed.
The projections of the fit on the $M_{bc}$ variable for $B^+\to \eta K^+$ and $B^-\to \eta K^-$ 
candidates, selected after further enhancing S/B with the requirements $-0.1<\Delta E<0.08$ GeV and $R^\prime>1.95$, are shown in Figure~\ref{fig:Belle:etah}.

\begin{table}[t]
\begin{center}
\caption{Results of the $B\to \eta h$ measurements by the Belle Collaboration~\cite{bib:Belle:etah}.}
\begin{tabular}{|l|c|c|c|c|}
\hline \textbf{Decay} & \textbf{BF $(10^{-6})$} & \textbf{BF significance $(\sigma)$} & \textbf{$A_{\CP}$} & \textbf{$A_{\CP}$ significance $(\sigma)$} \\
\hline $B^+\to \eta K^+$ & $2.12\pm 0.23 \pm 0.11$ & 13.2 & $-0.38 \pm 0.11 \pm 0.01$ & 3.8 \\
\hline $B^+\to \eta \pi^+$ & $4.07\pm 0.26 \pm 0.21$ & 22.4 & $-0.19\pm 0.06 \pm 0.01$ & 3.0 \\
\hline $B^0\to \eta K^0$ & $1.27^{+0.33}_{-0.29}\pm 0.08$ & 5.4 & -- & --  \\
\hline
\end{tabular}
\label{tab:Belle:etah}
\end{center}
\end{table}

The main results are summarized in Table~\ref{tab:Belle:etah}.
Both $B^+\to \eta K^+$ and $B^+\to \eta \pi^+$ signals are observed with large statistical 
significances, exceeding $10\sigma$ and $20\sigma$ respectively, and both $B$ decays 
exhibit large negative \CP asymmetries, around $-40\%$ and $-20\%$ respectively, at least
$3\sigma$ different from zero. The $B\to \eta K^0$ decay is observed for the first time, with a
significance of more than $5\sigma$, and its branching fraction is smaller than that of
$B^+\to \eta K^+$. All results are in agreement with SM expectations. 

\subsection{Search for \CP violation in three-body $B\to KKK$ decays}
Measurements of time-dependent \CP violation in $b\to q\bar q s$ $(q =
u, d, s)$ decays offer a method for determining $\beta$ alternative to the one 
based on $B\to c\bar c K^{(*)0}$.
Such decays are dominated by $b \to s$ loop diagrams, and therefore
are sensitive to possible new physics contributions appearing in
the loops of these diagrams. As a result, 
the effective $\beta$ ($\beta_{\rm eff}$) measured in such decays
could differ from $\beta$ measured in $B^0 \to c\bar c K^0$. 
% Deviations of $\beta_{\rm eff}$ from $\beta$ are also
% possible in the SM, due to additional amplitudes from $b\to u$ tree
% diagrams, loop diagrams containing different CKM factors (``$u$
% penguins''), and final-state interactions.
Two final states, the \CP-even $K^0_S K^0_S K^0_S$ and the \CP-odd $\phi K_S^0$, are 
particularly suited for a NP search, as $\beta_{\rm eff}$ 
for these decays is expected to be very close, within a few \%, to $\beta$ measured in $B$ decays to charmonium~\cite{bib:sin2betaeff1,bib:sin2betaeff2,bib:sin2betaeff3}, with very little theoretical
uncertainties.
The related decay mode $B^\pm \to \phi K^\pm$ is also a powerful probe to NP, since
it is dominated in the SM by a single $b\to s$ penguin amplitude, and its direct \CP asymmetry 
is predicted to be small, $A_{\CP} \approx (0.0-4.7)\%$.

\babar\ has published in~\cite{bib:BaBar:KsKsKs} a detailed study of $B\to K^0_SK^0_SK^0_S$
decays, including a time-integrated, \CP-averaged Dalitz plot analysis -- to extract the amplitude and phases of the various resonant and non-resonant contributions to the decay -- and a
measurement of $\sin 2\beta_{\rm eff}$ from the time-dependent \CP asymmetry.
In the Dalitz plot analysis, events are selected in which three $K^0_S$ decay to charged pion pairs,
have a common vertex and form a $B$ meson candidate with $\mes \approx m_B$ and 
$\Delta E \approx 0$. Continuum $q\bar{q}$ background is suppressed by means of a neural
network $NN$ of four event-shape variables. The $B^0\to K^0_SK^0_SK^0_S$ decay amplitude is 
modeled as the sum of a non-resonant term and several intermediate two-body amplitudes; 
the magnitudes and phases of each term are determined, together with the signal and background
yields, through a maximum likelihood fit to 
$\mes$, $\Delta E$, $NN$, and to the two Dalitz plot variables, $s_{\rm min}$ and $s_{\rm max}$,
which are the minimum and the maximum, respectively, among the three Mandelstam variables 
$s_{ij}$, where the indices $i, j=1..3$ ($i\ne j$) denote the three $K^0_S$.
Evidence is found for contributions to the final state by $f_0(980)K^0_S$, $f_0(1710)K^0_S$,
$f_2(2010)K^0_S$, $\chi_{c0}K^0_S$ and non-resonant $K^0_SK^0_SK^0_S$, while the fit
likelihood does not improve significantly when including in the model additional resonances, like
the broad $f_X(1500)$ used to explain some older data~\cite{bib:fX}.
The global maximum of the likelihood favors large non-resonant and small resonant 
contributions, with significant destructive interference among them.
The signal yield obtained from the fit is $200\pm 15$, while the efficiency, obtained from a simulation
of signal events where the magnitudes and phases of the intermediate resonances are fixed to the
central values returned by the fit, is $\varepsilon = 6.6\%$, yielding an inclusive branching fraction $BF(B^0\to K^0_SK^0_SK^0_S) = (6.19\pm 0.48 \pm 0.15\pm 0.12)\times 10^{-6}$.

\begin{figure*}[!htbp]
\centering
\includegraphics[width=0.48\textwidth]{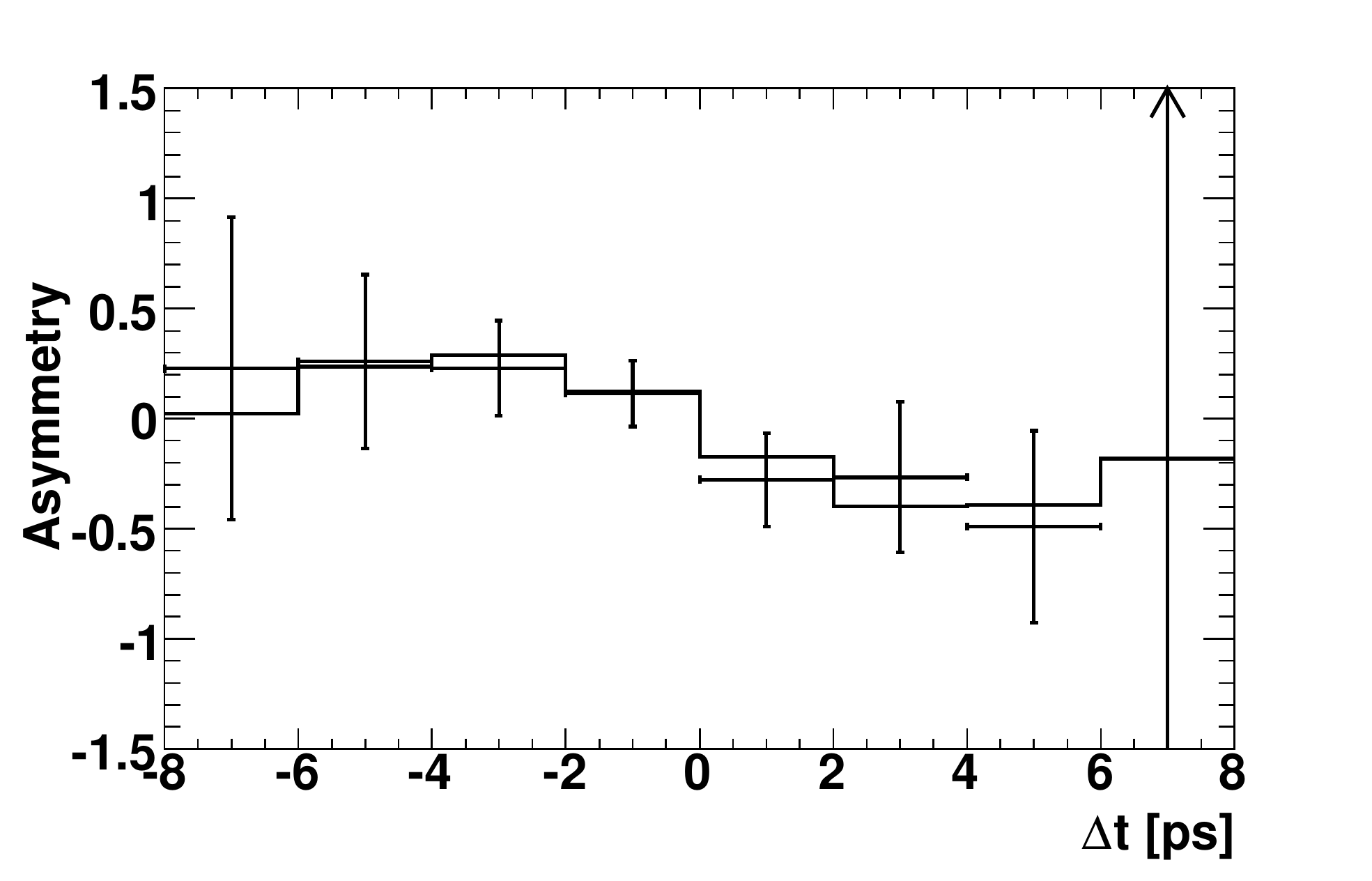}
\includegraphics[width=0.48\textwidth]{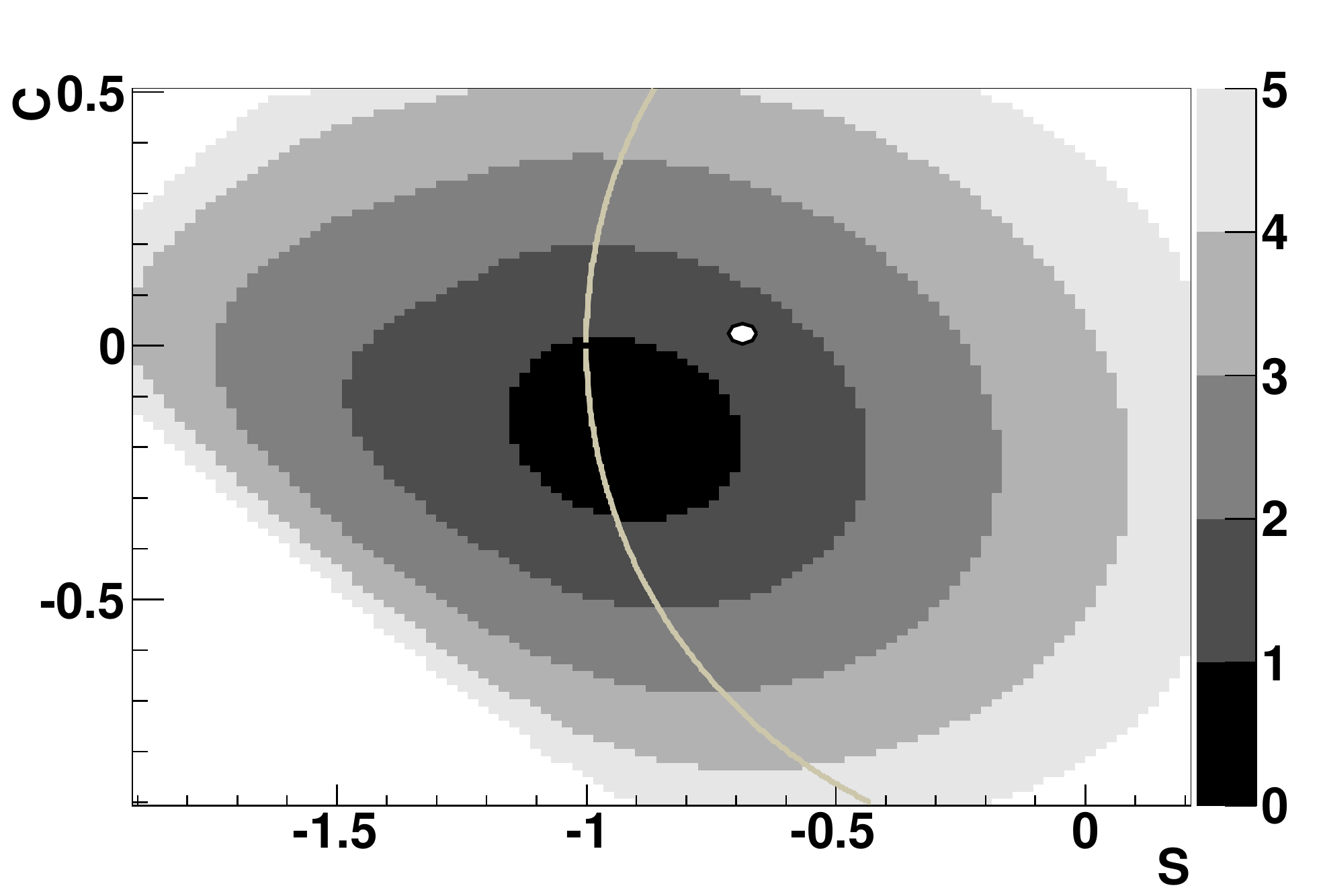}
\caption{
Left: Background-subtracted data distribution (points with error bars) and fit (histogram)
of the time-dependent \CP asymmetry for $B\to K^0_SK^0_SK^0_S$ candidates
reconstructed by \babar~\cite{bib:BaBar:KsKsKs}.
Right: two-dimensional scan of $-2\Delta ln \mathcal{L}$, including systematic uncertainties, 
as a function of $\mathcal{S}$ and $\mathcal{C}$.  The result of the \babar\ analysis of 
$B^0 \to c\bar{c} K^{(*)}$ decays is indicated as a white ellipse and the physical boundary $(\mathcal{S}^2+\mathcal{C}^2\le 1)$ is marked as a gray line.} 
\label{fig:BaBar:KsKsKs}
\end{figure*}

In the \CP violation analysis, events are selected in which one neutral $B_{\rm tag}$ decays to a (partially reconstructed) flavor-specific final state, thus
identifying the flavor of the other $B$ ($B_{\rm sig}$) as $B^0$ or $\overline{B}^0$ at the same time, while $B_{\rm sig}$ decays to either three $K_S^0 \to \pi^+\pi^-$ ($\varepsilon = 6.6\%$) or two $K_S^0 \to \pi^+\pi^-$ and one $K_S^0 \to \pi^0\pi^0$ ($\varepsilon = 3.1\%$).
The \CP asymmetry between the $B^0$ and $\overline{B}^0$ decays to
three $K^0_S$ has a dependence on the proper time difference $\Delta t$ between the two 
$B$ decays of the form 
$[\mathcal{S}\sin(\Delta m_d \Delta t) -\mathcal{C}\cos(\Delta m_d\Delta t)]$, 
where $\Delta m_d$ is the mass difference between the two neutral $B$ physical eigenstates.
In the SM $\mathcal{S} = -\sin 2\beta$, while $\mathcal{C}\approx 0$, since the decay amplitude is dominated by a single weak phase term and direct \CP violation is thus negligible.
$\mathcal{S}$ and $\mathcal{C}$ are extracted from a maximum likelihood fit to $\Delta t$, \mes, $\Delta E$ and $NN$ for the selected candidates (see Figure~\ref{fig:BaBar:KsKsKs}). 
The signal yield is $263\pm 21$, and the measured values of $\mathcal{S}$ and $\mathcal{C}$ are:
\begin{eqnarray} 
\mathcal{S} &=& -0.94 ^{+0.24}_{-0.21} ({\rm stat}) \pm 0.06 ({\rm syst}), \\
\mathcal{C} &=& -0.17 \pm 0.18 ({\rm stat}) \pm 0.24 ({\rm syst}),
\end{eqnarray}
where the systematic errors are dominated by the $\Delta t$ resolution and the fit bias.
They are compatible within two standard deviations with those measured in
tree-dominated modes such as $B^0 \to J/\psi K_S^0$, as expected
in the SM. 
\CP conservation ($\mathcal{S}= \mathcal{C}=0$) is excluded at $3.8\sigma$, including systematic 
uncertainties, thus providing for the first time evidence of \CP violation in $B^0 \to K^0K^0K^0$ 
decays.

\begin{figure*}[!htbp]
\centering
\includegraphics[width=0.31\textwidth]{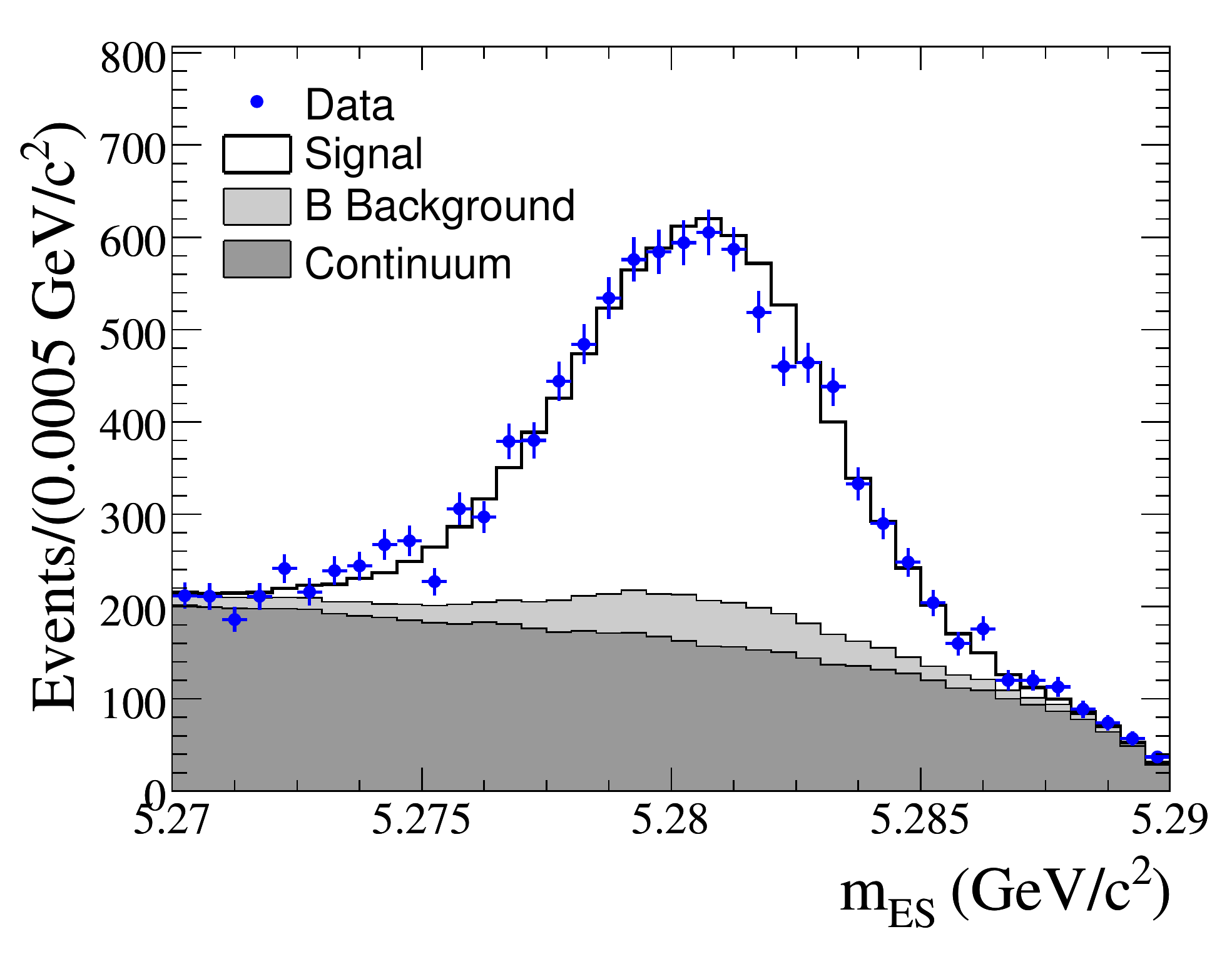}
\includegraphics[width=0.31\textwidth]{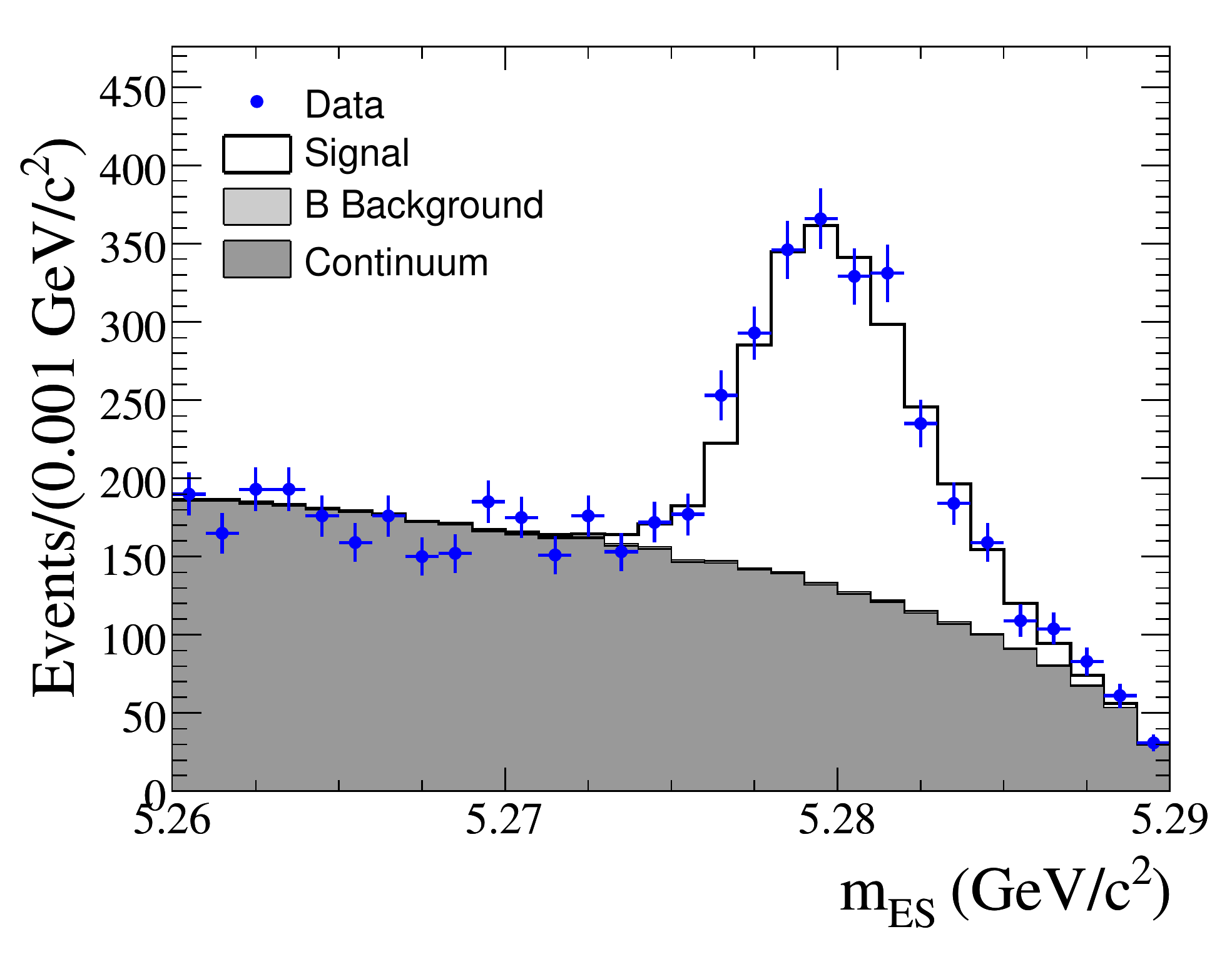}
\includegraphics[width=0.31\textwidth]{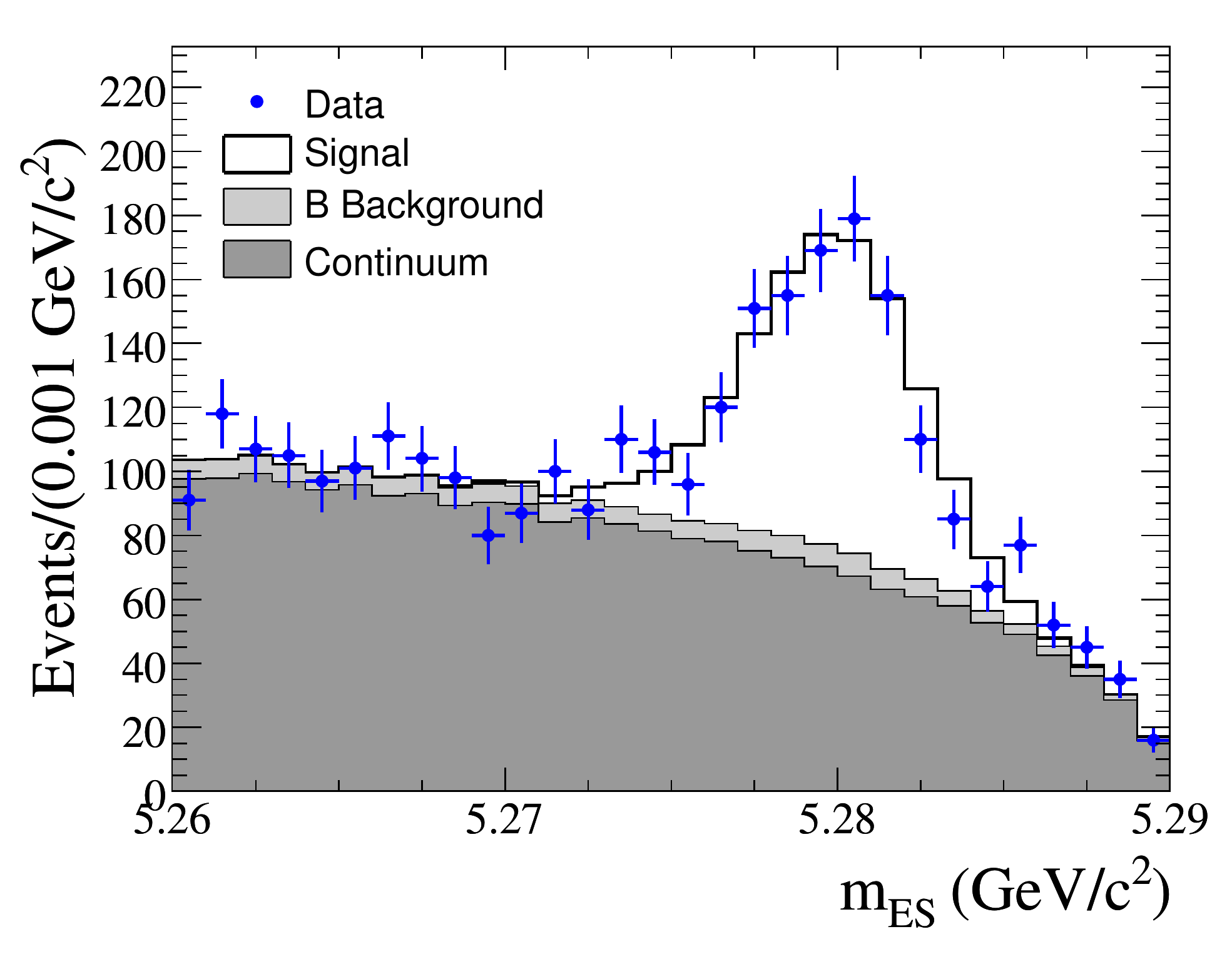}
\caption{Distributions of \mes for $B^+\to K^+ K^-K^+$ (left), $B^0\to K^+K^-K^0_S$, and $B^+\to K^0_SK^0_SK^+$ (right) selected by \babar~\cite{bib:BaBar:KKK} (points with error bars). The 
fit results are overlaid.} 
\label{fig:BaBar:KKK1}
\end{figure*}

\begin{figure*}[!htbp]
\centering
\includegraphics[width=0.31\textwidth]{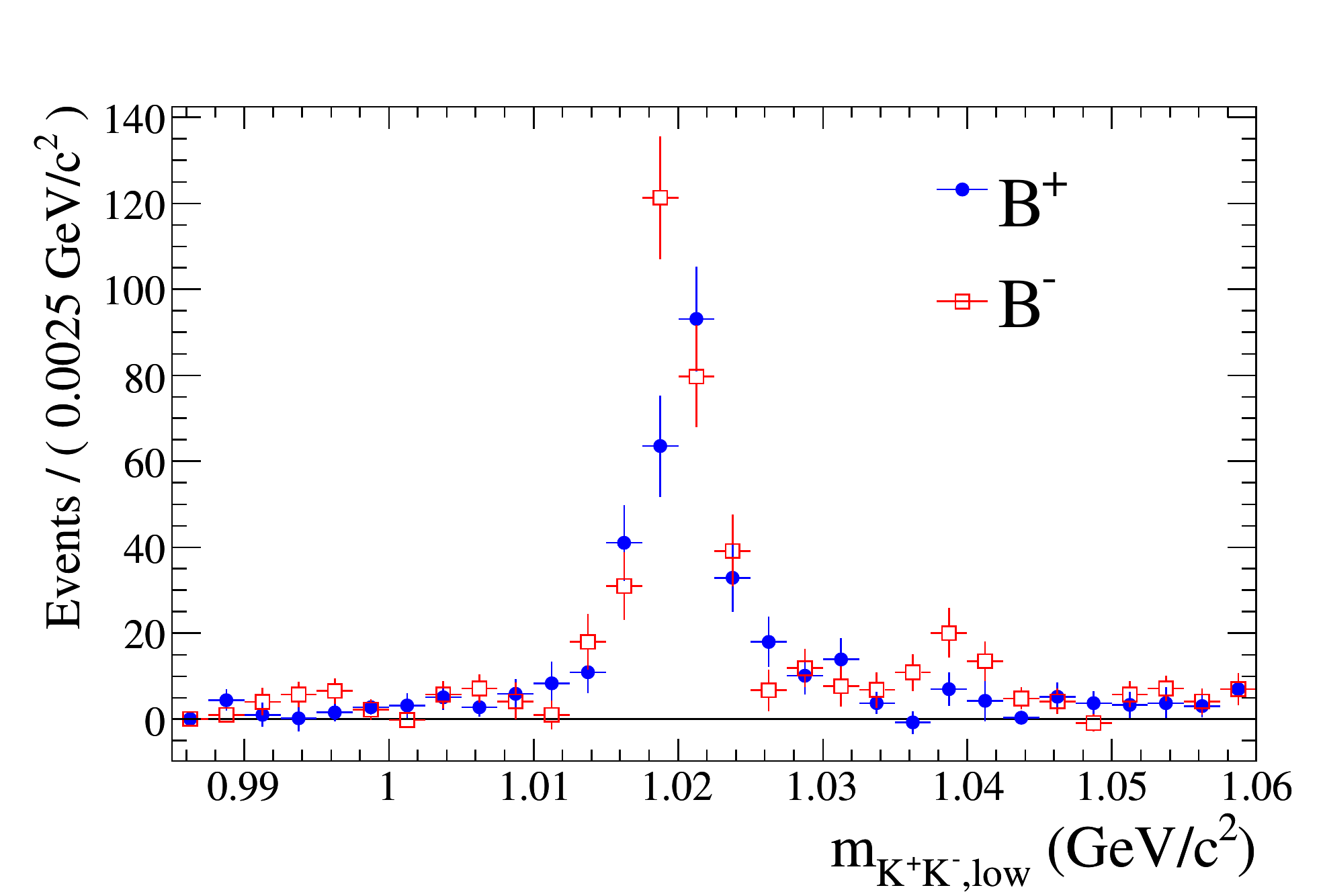}
\includegraphics[width=0.31\textwidth]{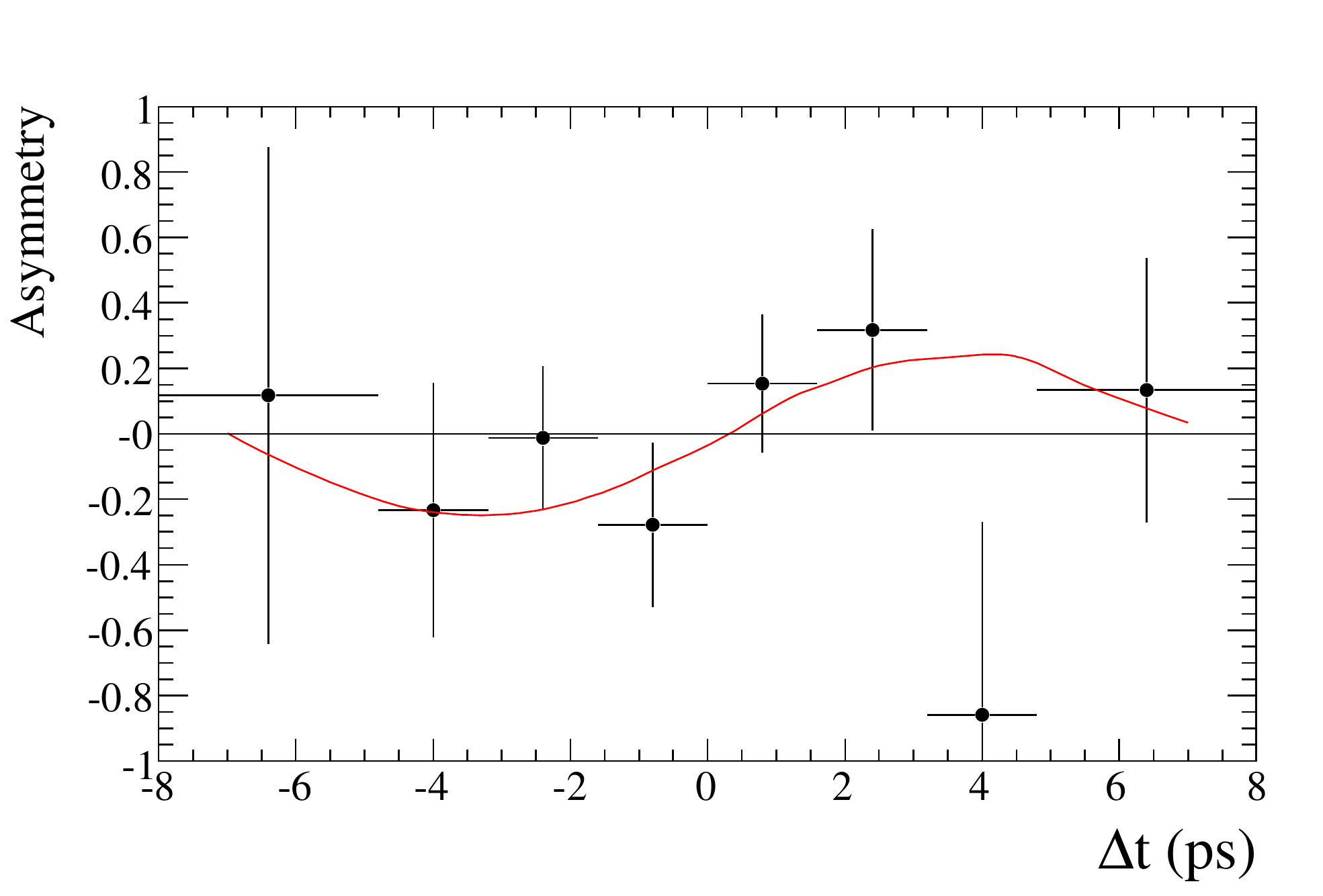}
\includegraphics[width=0.31\textwidth]{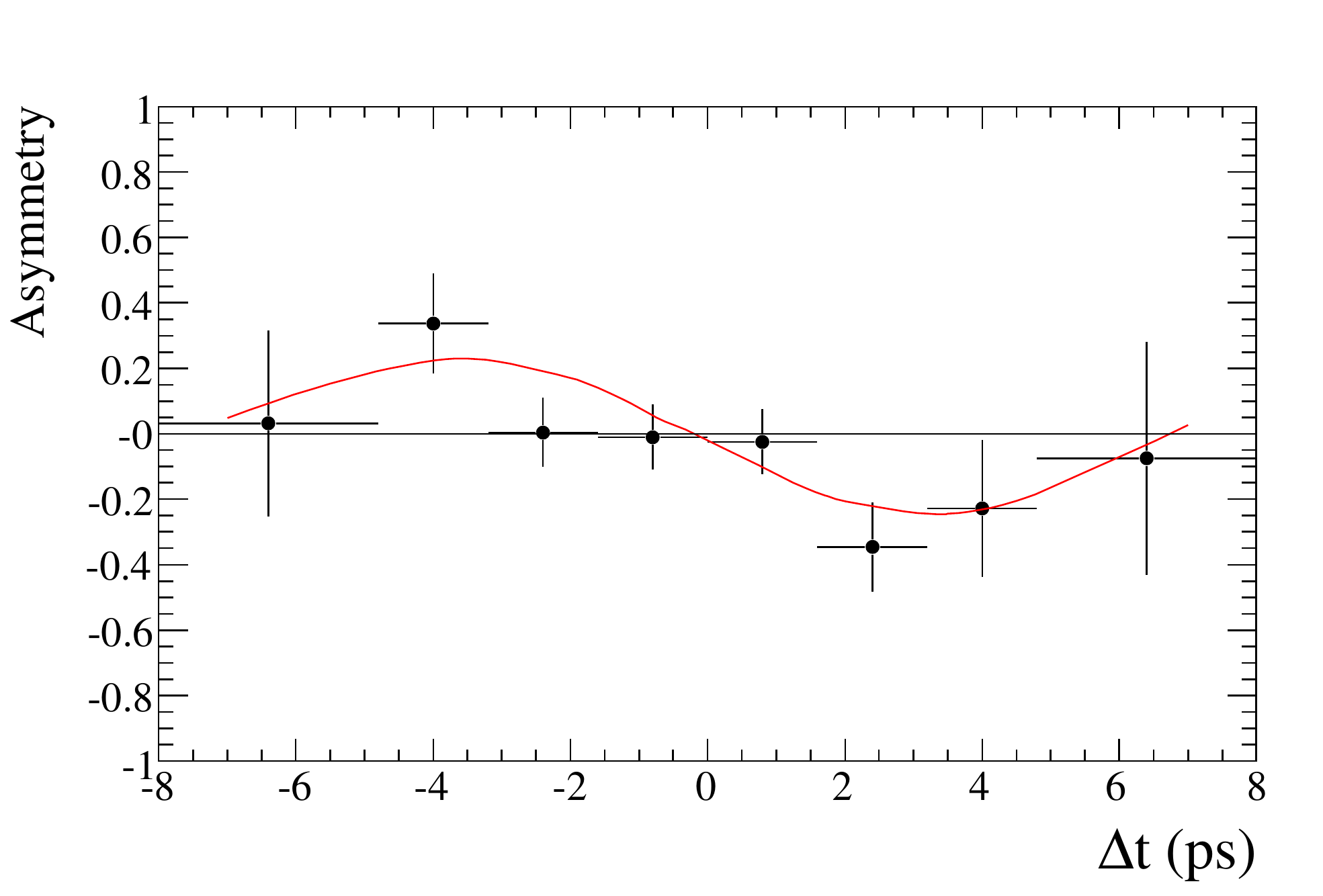}
\caption{Left: Background-subtracted distributions of $B^\pm \to K^+K^-K^\pm$ candidates in data, plotted separately for $B^+$ and $B^-$ events, in the ``$\phi(1020)$ region"  where $m_{K^+K^-}\approx m_\phi$ (considering the kaon pair with lowest invariant mass).
Center and right: time-dependent \CP asymmetry, as a function of $\Delta t$, for $B^0\to K^+K^-K^0_S$ candidates in the $\phi(1020)$ region (center) and in the $\phi(1020)$-excluded region (right). The points represent signal-weighted data, and the line is the fit model.
} 
\label{fig:BaBar:KKK2}
\end{figure*}

\babar\ has also published in~\cite{bib:BaBar:KKK} a detailed study of the other $B$ decays
to three kaons. The decay $B^0\to K^+K^- K^0_S$ is used, by performing a 
time-dependent analysis, to measure $\beta_{\rm eff}$ in $B^0$ decays to the
\CP-odd final state $\phi K^0_S$ and also to the (mostly \CP-even) other resonant and
non-resonant $K^+K^-K^0_S$ final states. 
The decay $B^0\to K^+K^- K^+$ is used to determine the direct \CP asymmetry in 
$B^+ \to \phi K^+$. In order to disentangle the $B\to \phi K$ contribution from other 
resonant or non-resonant contributions, a Dalitz plot analysis is performed for each of the 
final states considered. In addition, to better understand the Dalitz plot structure of these
decays and the possible intermediate $KK$ resonances, also $B^+\to K^0_SK^0_SK^+$
decays are studied, since they have a simplified spin structure due to the fact that the two $K_S^0$
mesons in the final state are forbidden (by Bose--Einstein statistics) to be in an odd angular 
momentum configuration.
Events with three charged or neutral kaons kinematically consistent with a $B$ decay (common
production vertex, $\mes\approx m_B$, $\Delta E\approx 0$) are selected. Continuum background
events are suppressed by means of a neural network $NN$ of typically five event-shape variables.
The signal $B$ decay amplitude is modeled as the sum of a non-resonant and of several
resonant contributions, whose magnitudes and phases are determined through a maximum
likelihood fit to the Dalitz plot variables, together with $\mes$, $\Delta E$ and $NN$. Only 
resonances that significantly improve the quality of the fit are kept in the nominal signal decay
amplitude: in all the three channels no evidence is found for the broad $f_X(1500)$.
The selection efficiency is around 30\% for each of the three decay modes.
The number of selected signal decays, the measured inclusive charmless branching fractions 
(excluding the $\chi_{c0}K$ contributions) and the 
\CP-violating parameters are summarized in Table~\ref{tab:BaBar:KKK}. Some of the fit results are
also illustrated in Figures~\ref{fig:BaBar:KKK1} and~\ref{fig:BaBar:KKK2}.
The \CP asymmetry in $B^+\to\phi K^+$ differs from 0 by $2.8\sigma$, and is in slight tension with
SM predictions ($0-4.7\%$). The amount of \CP violation in $B^0\to \phi K^0_S$ and in other
$B^0\to K^+K^-K^0_S$ decays is in excellent agreement with SM expectations: $\beta_{\rm eff}$ is 
very close to $\beta$ measured in $B$ decays to charmonium, and $\mathcal{C}$ is consistent with
zero.

\begin{table}[!htbp]
  \caption{Results of the $B \to KKK$ measurements by the \babar\ 
  Collaboration~\cite{bib:BaBar:KKK}. For neutral $B$ decays, the \CP asymmetry $A_{\CP}$ is
  defined as $A_{\CP} = -\mathcal{C}$.}
    \label{tab:BaBar:KKK}
\begin{tabular}{|l|c|c|c|c|}
\hline
\textbf{Decay} & \textbf{$N_{\rm sig}$} & \textbf{BF $(10^{-6})$} & \textbf{$A_{\CP}$} & 
$\beta_{\rm eff}$\\
\hline
\multirow{2}{*}{$B^+\to K^+K^-K^+$} & \multirow{2}{*}{$5269\pm 84$} & \multirow{2}{*}{$33.4 \pm 0.5 \pm 0.9$} & $(-1.7^{+1.9}_{-1.4} \pm 1.4)\%$ (inclusive) & \multirow{2}{*}{$-$} \\
&  &  & $(12.8 \pm 4.4 \pm 1.3)\%$ ($\phi K^+$) & \\
\hline
\multirow{3}{*}{$B^0\to K^+ K^- K^0_S$} & \multirow{3}{*}{$1579\pm 46$} & \multirow{3}{*}{$25.4\pm 0.9 \pm 0.8$} & $(-5 \pm 18\pm 5)\%$ ($\phi K^0_S$) & $(21\pm 6\pm 2)^\circ$ ($\phi K^0_S$) \\
& & & $(-28\pm 24 \pm 9)^\circ$ ($f_0(980)K^0_S$)& $(18\pm 6 \pm 4)^\circ$ ($f_0(980)K^0_S$) \\
& & & $(-2\pm 9 \pm 3)^\circ$ (Other)& $(20.3\pm 4.3 \pm 1.2)^\circ$ (Other) \\
\hline
$B^+ \to K^0_S K^0_S K^+$ & $632\pm 28$ & $10.1 \pm 0.5 \pm 0.3$ & $(4\pm 5 \pm 2)\%$ & $-$ \\
\hline
\end{tabular}
\end{table}

\subsection{Search for \CP violation in three-body $B\to K\pi^0\pi^0$ decays}
Recent measurements of rates and asymmetries in $B\to K\pi$ decays may hint 
to possible new physics contributions, but hadronic uncertainties prevent a clear
conclusion in this direction. 
Additional information can be obtained through a data-driven approach involving 
measurements of all the observables in the $B \to K\pi$ system, or looking at the related 
pseudoscalar-vector decays $B \to K^*\pi$ and $B\to K\rho$. In such decays, 
the ratios of tree-to-penguin amplitudes are predicted to be 2 to 3 times larger than those in 
$B\to K\pi$, possibly leading to larger \CP asymmetries~\cite{bib:theo:Kpi0pi01,bib:theo:Kpi0pi02,bib:theo:Kpi0pi03}.

\babar\ has published in~\cite{bib:BaBar:Kpi0pi0} a measurement of the branching
fraction and \CP asymmetry of the decay $B^\pm\to K^{*}(892)^\pm\pi^0$, based on the final
$\Upsilon(4S)$ sample. The $K^*(892)^+$ is reconstructed in the $K^+\pi^0$ decay
channel, and discrimination between $B^+\to K^*(892)^+\pi^0$ decays and other
(resonant or non-resonant) $B^+\to K^+\pi^0\pi^0$ decays is achieved through inspection
of the two-body invariant mass distributions of the particles in the final state, 
since the Dalitz plot of this decay is characterized by rather narrow resonances ($K^*(892)^+$, 
$f_0(980)$, $\chi_{c0}$). 
\begin{figure*}[!htbp]
\centering
\includegraphics[width=0.48\textwidth]{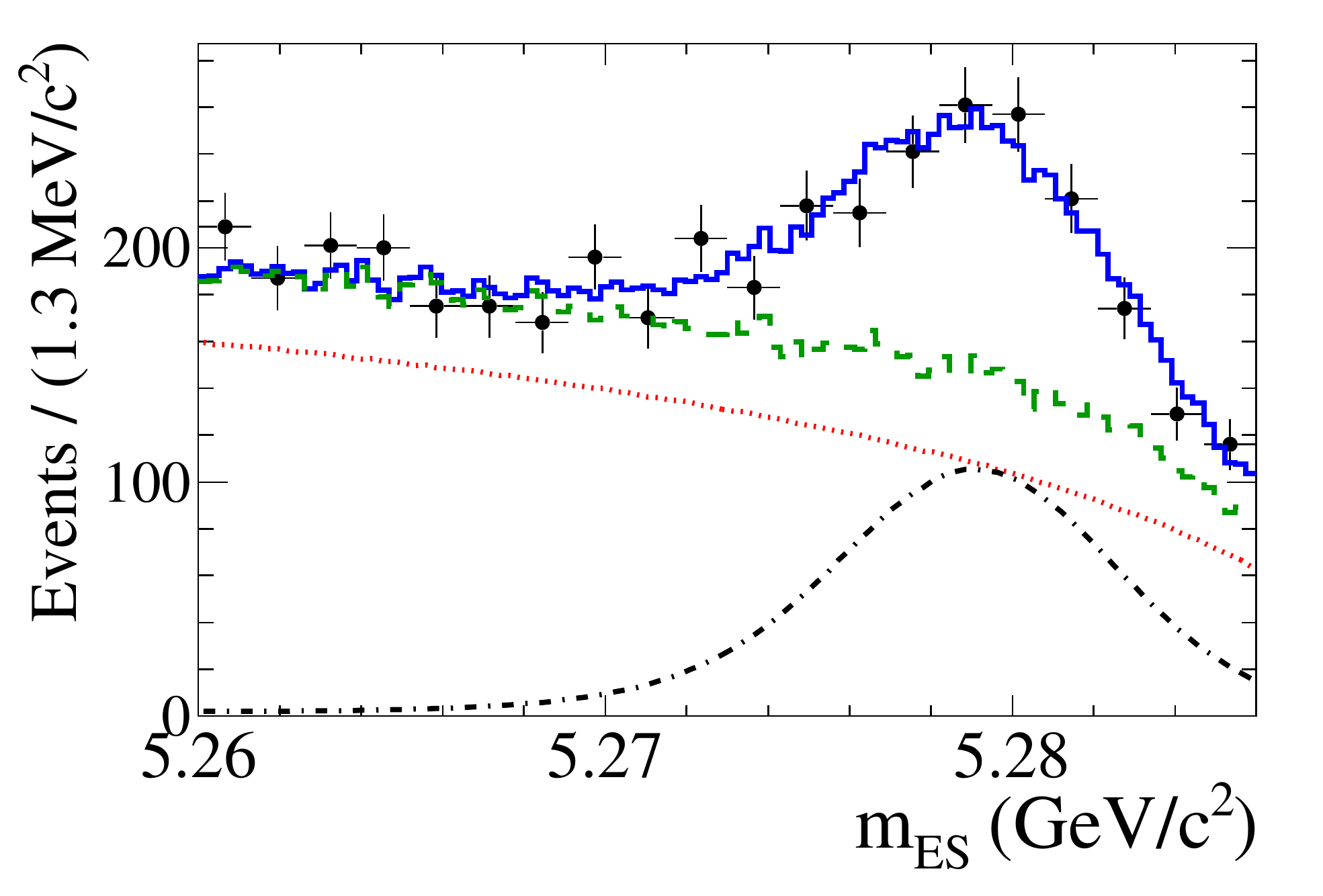}
\includegraphics[width=0.48\textwidth]{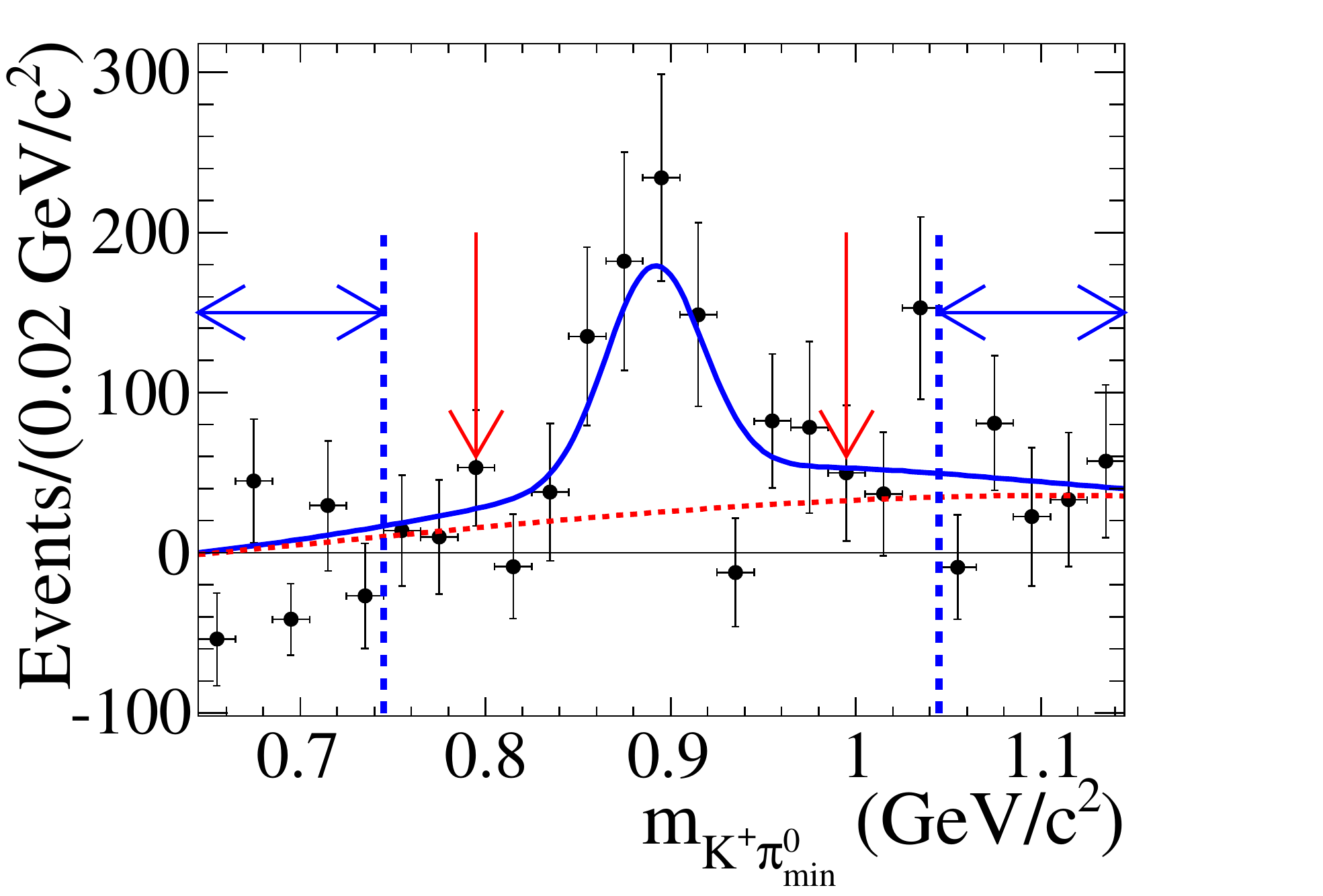}
\caption{Left: \mes distribution of $B^\pm\to K^\pm\pi^0\pi^0$ candidates selected in data
by \babar~\cite{bib:BaBar:Kpi0pi0}, after applying a requirement on the neural-network variable
that retains 60\% of the signal while suppressing a large fraction of the $q\bar{q}$ 
background. The solid (blue) lines represents the \mes projection of the total fit, 
the dashed (green) lines the total background contribution, and the dotted (red) lines the 
$q\bar{q}$ component. The dash-dotted lines represent the signal contribution.
Right: efficiency-corrected, background-subtracted distribution of the invariant mass
$m_{K^+\pi^0_{\rm min}}$ for $B^\pm\to K^\pm\pi^0\pi^0$ candidates where 
$m_{K^+\pi^0_{\rm min}}$ is close to the $K^*(892)^+$ mass.
$m_{{K^+\pi^0}_{\rm min}}$ is the mass of the $K^+\pi^0$ combination with lower
invariant mass. The vertical red arrows define the limits of the signal region while the blue horizontal
arrows define the sideband background control regions.The curves show the results of the fit
used to cross-check the procedure, for the total (blue solid line) and background-only (red 
dashed line) components.
} 
\label{fig:BaBar:Kpi0pi0}
\end{figure*}
A total $B^+\to K^+\pi^0\pi^0$ yield of $1220\pm 85$ events (significance $>10\sigma$) is 
obtained from a ML fit to \mes and to a neural network $NN$ of event-shape variables 
(see Figure~\ref{fig:BaBar:Kpi0pi0}, left).
The contribution from $B^+\to K^{*}(892)^+\pi^0$ is obtained by subtracting non-$B^+\to K^+\pi^0\pi^0$ events using the $sPlot$ technique and then studying the distribution of $m_{K^+\pi^0_{\rm min}}$,
the invariant mass of the $K^+\pi^0$ combination with lower invariant mass. The sidebands of 
$m_{K^+\pi^0_{\rm min}}$ are used to estimate the non-$K^*(892)^+\pi^0$ background in the
signal region near the $K^*(892)^+$ mass, which is then subtracted from the number of 
candidates observed in the signal $m_{K^+\pi^0_{\rm min}}$ region (see 
Figure~\ref{fig:BaBar:Kpi0pi0}, right).
The final results are summarized in Table~\ref{tab:BaBar:Kpi0pi0}. For 
$B^\pm \to K^*(892)^\pm\pi^0$, a branching fraction of $(8.2\pm 1.5 \pm 1.1)\times 10^{-6}$
is measured, with a \CP asymmetry close to zero ($-6\%$) but also consistent, within errors, 
with large \CP violation ($A_{\CP} \approx 20-30\%$) as expected in the SM.

\begin{table}[t]
  \caption{
    Summary of measurements of branching fractions (averaged over charge
    conjugate states) and \CP asymmetries measured by \babar\ in 
    $B^+\to K^+\pi^0\pi^0$ decays~\cite{bib:BaBar:Kpi0pi0}. 
    Both product branching
    fractions and those corrected for secondary decays are shown.  For each
    result, the first uncertainty is statistical, the second is
    systematic and the third, where quoted, is the error on
    $\chi_{c0}\to\pi^0\pi^0$. The notation $Rh$ refers, where applicable, to the
    intermediate state of a resonance and a bachelor hadron. }
    \label{tab:BaBar:Kpi0pi0}
\begin{tabular}{|l|c|c|c|}
\hline
\textbf{Decay} & \textbf{$BF(B^+ \to Rh \to K^+\pi^0\pi^0) (10^{-6})$} & \textbf{$BF(B^+ \to Rh) (10^{-6})$} & \textbf{$A_{\CP}$} \\
\hline
$B^+ \to K^+ \pi^0 \pi^0$ &$16.2 \pm 1.2 \pm 1.5$  &  $-$   & $-0.06\pm0.06\pm0.04$ \\
\hline
$B^+ \to K^*(892) \pi^0$ & $2.7 \pm 0.5\pm 0.4$     & $8.2\pm 1.5\pm1.1$    & $-0.06\pm 0.24 \pm 0.04$ \\
$B^+ \to f_0(980) K^+$   & $2.8 \pm 0.6 \pm 0.5$     & $-$   & $\phantom{-}0.18\pm 0.18\pm 0.04$ \\
$B^+ \to \chi_{c0} K^+$   & $0.51\pm 0.22 \pm 0.09$ & $180\pm 80 \pm 30 \pm 10$  & $-0.96 \pm 0.37 \pm 0.04$ \\
\hline
\end{tabular}
\end{table}

\section{SEARCH FOR NEW PHYSICS IN VERY RARE $B$ DECAYS}
Some $B$ meson decays are expected in the Standard Model to have branching fractions 
below the experimental sensitivity of the $B$-factories. The observation of such very rare 
decays would therefore provide evidence for new physics beyond the SM.
This is the case for instance of lepton-flavor violating $B$ decays and leptonic 
$B\to \nu\bar\nu$ decays.

\subsection{Search for lepton flavor violation in $B\to h\tau \ell$}
Lepton flavor violating decays of $B$ mesons can occur in the SM through loop processes that 
involve neutrino mixing. These are highly suppressed by powers of $m^2_\nu/m^2_W$, and have 
predicted branching fractions many orders of magnitude below the current experimental sensitivity. 
However, in many extensions of the SM, $B$ decays involving lepton flavor violation are greatly 
enhanced. In some cases, decays involving the second and third generations of quarks and leptons 
are particularly sensitive to physics beyond the 
SM~\cite{bib:theory:htaul1,bib:theory:htaul2,bib:theory:htaul3,bib:theory:htaul4}.

\begin{figure*}[!htbp]
\centering
\includegraphics[width=0.48\textwidth]{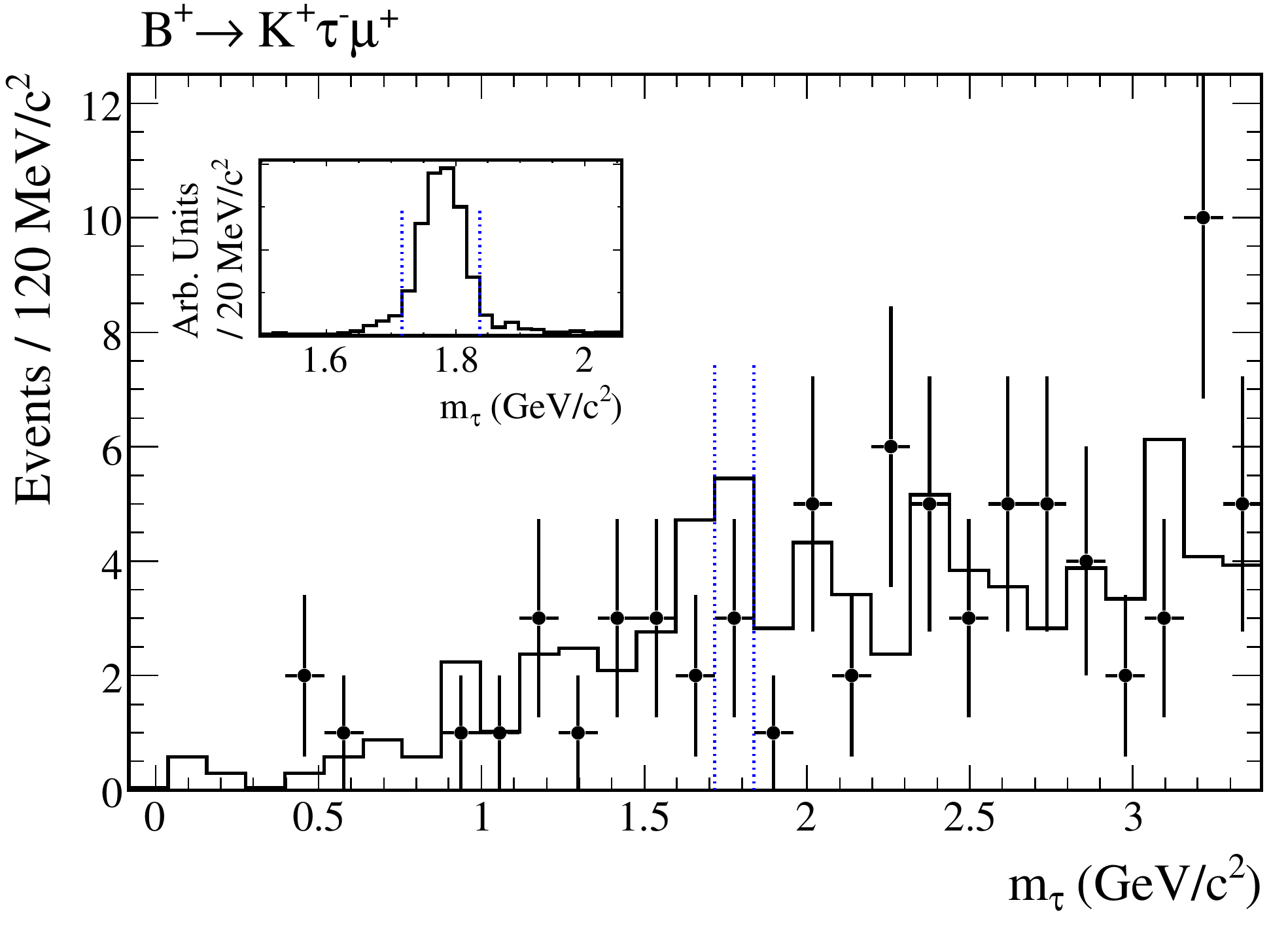}
\includegraphics[width=0.48\textwidth]{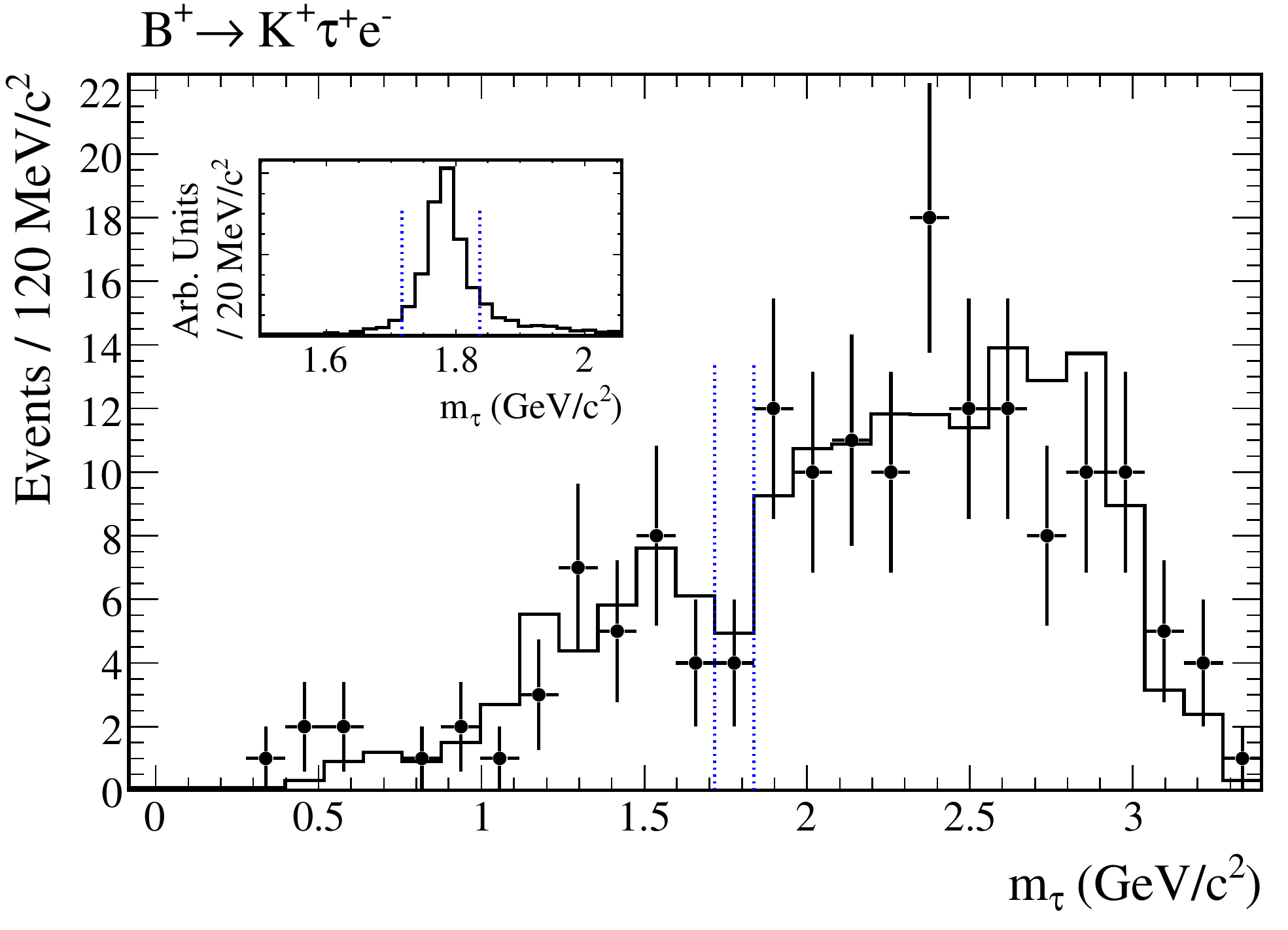}
\caption{Observed distributions of the $\tau$ invariant mass for the $B^+\to K^+\tau^-\mu^+$ 
candidates (left) and the $B^+\to K^+\tau^+ e^-$ candidates (right) selected by
\babar~\cite{bib:BaBar:htaul}.
         The distributions show the sum of the three $\tau$ channels ($e$, $\mu$, $\pi$).
         The points with error bars are the data.
         The solid line is the background MC which has been normalized to the
         area of the data distribution.
         The dashed vertical lines indicate the $m_\tau$ signal window range.
         The inset shows the $m_\tau$ distribution for signal MC.} 
\label{fig:BaBar:htaul1}
\end{figure*}

\begin{figure*}[!htbp]
\centering
\includegraphics[width=0.48\textwidth]{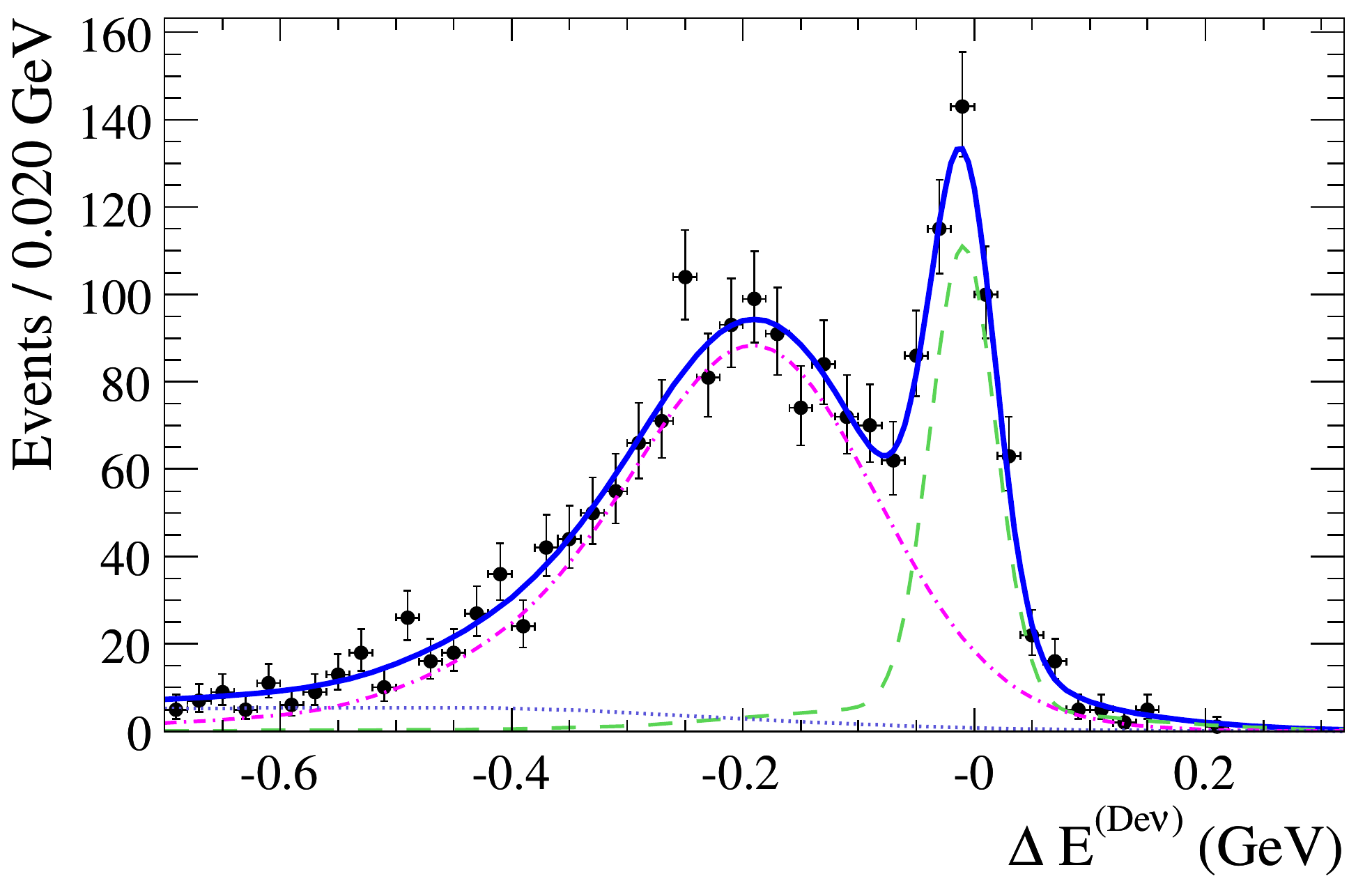}
\includegraphics[width=0.48\textwidth]{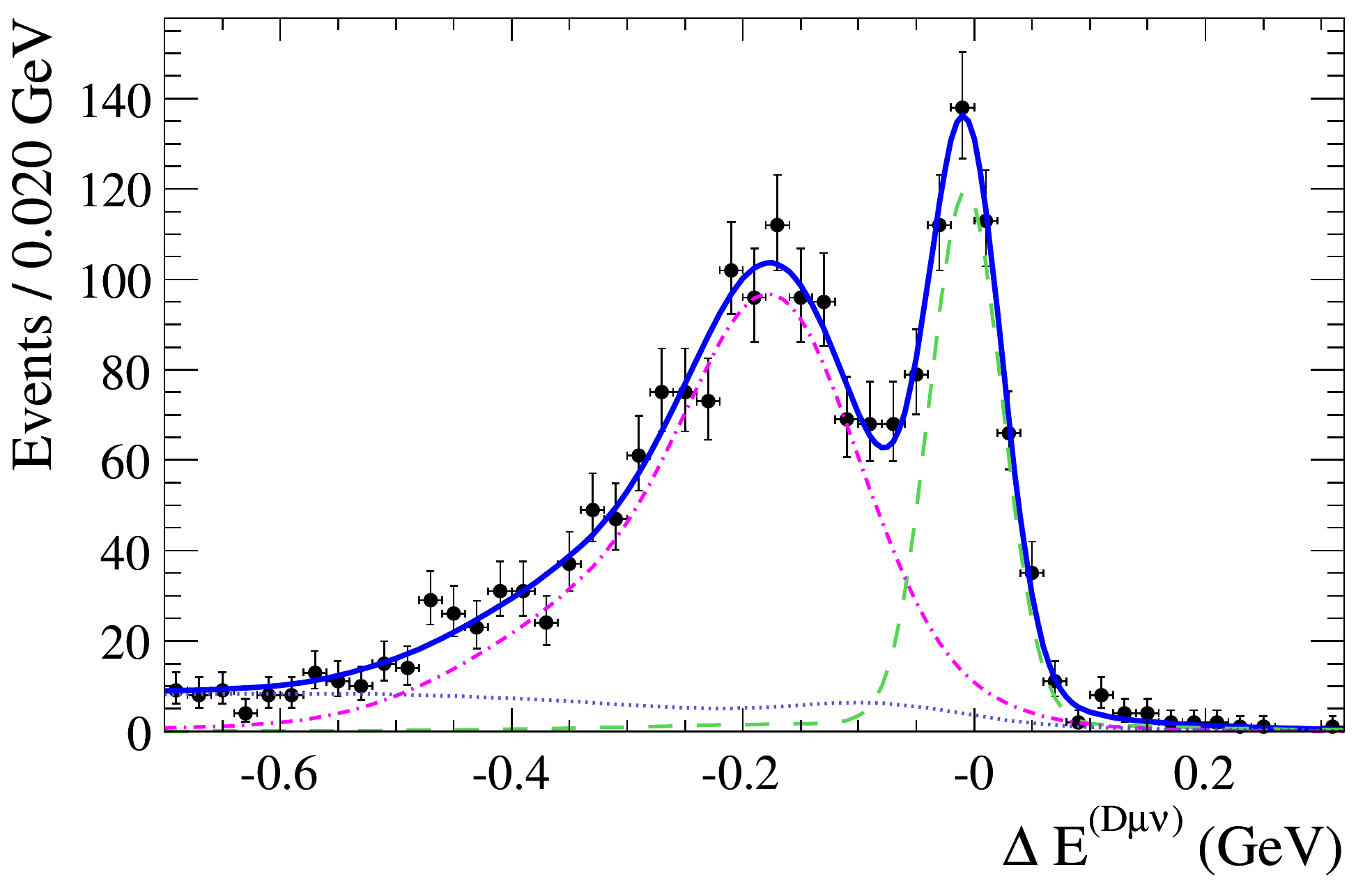}
\caption{Distributions of the three-component $\Delta E_{D\ell \nu}$ unbinned 
maximum likelihood fits of the data for the $B\to D^{(*)0}e\nu$ (left) and 
$B\to D^{(*)0}\mu\nu$ (right) control samples selected in the $B\to h\tau\ell$ \babar\
analysis~\cite{bib:BaBar:htaul}. In each plot, the points represent the data, the solid blue curve is the
fit result, the long-dashed green curve is the $D^0$ component, the dash-dotted purple curve is the $D^{*0}$ component, and the dotted blue curve is the $D^{**0}$ component, which also includes any residual combinatorial background.} 
\label{fig:BaBar:htaul2}
\end{figure*}

\babar\ has presented at the FPCP 2012 conference, and later published in~\cite{bib:BaBar:htaul},
a search for the eight lepton flavor violating decays $B^\pm\to h^\pm\tau \ell$ , $h=\pi, K$, 
$\ell=e,\mu$, using the final $\Upsilon(4S)$ dataset.
Events in which one $B$ meson ($B_{\rm tag}$) is fully reconstructed in one of several 
hadronic final states, $B_{\rm tag} \to D^{(*)0}~n_1\pi^\pm~n_2 K^\pm~n_3 K^0_S~n_4 \pi^0$, are 
selected.
The number of tracks from the signal $B$ ($B_{\rm sig}$) decay and their particle identification 
information are required to be consistent with that of a $B^\pm \to h^\pm \tau \ell$ decay
followed by a one-prong $\tau$ decay (producing either a charged electron, muon, or pion), 
in order to improve the signal-to-background ratio.
Continuum background is suppressed using a likelihood ratio based on event shape information, 
unassociated calorimeter clusters, and the quality of muon identification for channels that have a
muon in the final state. Background from semileptonic $B$ or $D$ decays is suppressed by 
rejecting events where two oppositely-charged $B_{\rm sig}$ daughters are found to be kinematically 
compatible with originating from a charm decay, by computing their invariant mass in the hypothesis
of a $K\pi$ pair.
Using the momenta of the reconstructed $B$, $h$, and $\ell$ candidates, the four momentum of
the $\tau$ lepton in the center-of-mass frame is fully determined, thus circumventing the problem of measuring
 the undetected neutrino(s) from the $\tau$ decay:
\begin{eqnarray}
\vec{p}^*_{\tau} & = & -\vec{p}^*_{\rm tag} -\vec{p}^*_h-\vec{p}^*_\ell \\
E_{\tau}^* & = & E^*_{\rm beam} - E^*_h - E^*_\ell
\end{eqnarray}
The resulting $\tau$ candidate mass, $m_\tau = \sqrt{{E_\tau^*}^2 - |\vec{p}^*_\tau |^2}$, is then used to 
discriminate against combinatorial background (see Figure~\ref{fig:BaBar:htaul1}).
A broad $m_\tau$ sideband ($[0,3.5]$ GeV/$c^2$) is used to estimate the background in the signal 
region of $m_\tau$ close to the PDG $\tau$ mass, using a simulation of background events to
relate the number of background candidates in the sideband region to the number of background 
candidates in the signal region.
To reduce systematic uncertainties, the signal branching fraction is determined by using the ratio of 
the number of $B\to h\tau \ell$ signal candidates to the yield of control samples of 
$B^- \to D^{(*)0}\ell^-\bar\nu, D^0\to K^-\pi^+$ decays in events with a fully reconstructed 
hadronic $B_{\rm tag}$:
\begin{equation}
BF_{h\tau \ell} = BF_{D\ell\nu} \frac{N_{h\tau \ell}}{N_{D\ell\nu}} \frac{\varepsilon^{\rm sig}_{D\ell\nu}}{\varepsilon^{\rm sig}_{h\tau \ell}} \frac{\varepsilon^{\rm tag}_{D\ell\nu}}{\varepsilon^{\rm tag}_{h\tau \ell}}.
\end{equation}
Here $\varepsilon^{\rm tag}_{h\tau\ell}$ and $\varepsilon^{\rm tag}_{D\ell\nu}$ are the 
(similar) $B_{\rm tag}$ reconstruction efficiencies for events containing the different $B_{\rm sig}$ 
decays, while $\varepsilon^{\rm sig}_{h\tau\ell}$ and $\varepsilon^{\rm sig}_{D\ell\nu}$ are the signal
selection efficiencies. A charged $B_{\rm tag}$ candidate is properly reconstructed in approximately 
0.25\% of all $B\overline{B}$ events. 
The $B\to h\tau \ell$ selection efficiency varies between 2\% and 10\%
depending on the final state; the $D^{(*)0}\ell \nu$ signal efficiencies are close to 50\%.
The $D^0\ell\nu$ and $D^{*0}\ell\nu$ yields are determined through a fit to the energy difference
of the selected candidates,
\begin{equation}
\Delta E_{D\ell\nu} = E^*_K + E^*_\pi + E^*_\ell + E^*_\nu - 
E^*_{\rm beam},
\end{equation}
 where $E^*_\nu = p^*_\nu = |-\vec{p}^*_{\rm tag} - \vec{p}^*_K - \vec{p}^*_\pi - 
\vec{p}^*_\ell|$. The energy difference is peaked at 0 for $B\to D^0 \ell \nu$ and is peaked at lower
values ($\approx -200$ MeV) for $B\to D^{*0} \ell \nu$ (because of the undetected $\gamma$ or
$\pi^0$ from the $D^{*0}$ decay), while $B\to D^{**0}\ell\nu$ and other backgrounds do not peak
in $\Delta E_{D\ell\nu}$ (see Figure~\ref{fig:BaBar:htaul2}).

No evidence for the $B\to h\tau\ell$ decays is found, and 90\% confidence level upper limits are 
set -- with a frequentist technique -- on each branching fraction at the level of a few times $10^{-5}$,
as summarized in Table~\ref{tab:BaBar:htaul}.
The results are used to improve model-independent limits on the energy scale of new physics
in flavor-changing sermonic effective operators~\cite{bib:theory:htaul3} to 
$\Lambda_{bd} > 11$ TeV and $\Lambda_{bs} > $15 TeV.

   \begin{table}[!htbp]
     \begin{center}
       \caption{
          Branching fraction central values and 90\% C.L. upper limits (UL)
          for 
          $BF(B^+ \to h^+ \tau \ell)$.
          The uncertainties include statistical and systematic sources.
       }
       \label{tab:BaBar:htaul}
       \begin{tabular}{|l|c|c|}
         \hline
                      &  \multicolumn{2}{|c|}{ $BF(B \to h \tau \ell)$ $(\times 10^{-5})$} \\
         \multicolumn{1}{|c|}{ \ \ \ \ \  Mode \ \ \ \ \ \ \ \ \ \ \ }       &  \ \ \ \ central value\ \ \ \ & \ \ \ 90\% C.L. UL \ \ \ \\
         \hline
         $B^+ \to K^+ \tau^- \mu^+$   &  $\phantom{-}0.8\ ^{+1.9}_{-1.4}$   &  $<4.5$   \\
         \hline
         $B^+ \to K^+ \tau^+ \mu^-$   &  $-0.4\ ^{+1.4}_{-0.9}$   &  $<2.8$   \\
         \hline
         $B^+ \to K^+ \tau^- e^+$     &  $\phantom{-}0.2\ ^{+2.1}_{-1.0}$  &  $<4.3$   \\
         \hline
         $B^+ \to K^+ \tau^+ e^-$     &  $-1.3\ ^{+1.5}_{-1.8}$  &  $<1.5$   \\
         \hline
         $B^+ \to \pi^+ \tau^- \mu^+$ &  $\phantom{-}0.4\ ^{+3.1}_{-2.2}$  &  $<6.2$   \\
         \hline
         $B^+ \to \pi^+ \tau^+ \mu^-$ &  $\phantom{-}0.0\ ^{+2.6}_{-2.0}$  &  $<4.5$   \\
         \hline
         $B^+ \to \pi^+ \tau^- e^+$   &  $\phantom{-}2.8\ ^{+2.4}_{-1.9}$  &  $<7.4$   \\
         \hline
         $B^+ \to \pi^+ \tau^+ e^-$   &  $-3.1\ ^{+2.4}_{-2.1}$  &  $<2.0$   \\
         \hline
       \end{tabular}
     \end{center}
   \end{table}

\subsection{Search for new physics in $B^0$ decays to invisible final states and to $\nu\bar\nu\gamma$}
The decay $B^0 \to \nu \bar\nu$, which would give an invisible experimental signature, is strongly 
helicity-suppressed in the SM by a factor of order $(m_\nu/m_B)^2$ and its branching fraction is 
thus well below the $B$ factory experimental observability~\cite{bib:theo:invisible1}.
The SM expectation for the
$B^0 \to \nu\bar\nu\gamma$ branching fraction is predicted to be of order $10^{-9}$, with very little 
uncertainty from hadronic interactions~\cite{bib:theo:invisible2}. An experimental observation of an invisible $(+\gamma)$ 
decay of a $B^0$ at the $B$ factories would thus be a clear sign of physics beyond the SM.
Supersymmetric models or models with large extra dimensions allow significant, although small, 
rates for invisible $B^0$ decays, with branching fractions up to the $10^{-7}$ -- $10^{-6}$ 
range~\cite{bib:theo:invisible3,bib:theo:invisible4,bib:theo:invisible5}. 

Both Belle and \babar\ have presented at the FPCP 2012 conference preliminary results on 
invisible $B^0$ decays; \babar\ has also searched for $B^0 \to \nu\bar\nu\gamma$ decays.
The results have been later publicly documented in~\cite{bib:Belle:invisible} 
and~\cite{bib:BaBar:invisible}. 
The analysis technique is similar for the two experiments: only events where a neutral $B_{\rm tag}$
meson is found are considered, and the rest of the event (the ``signal side") is checked for consistency 
with an invisible decay or a decay to a single photon of the other neutral $B$.
Events with extra tracks or neutral pions or $K^0_L$ candidates are vetoed.
The total energy $E_{\rm extra}$ (or $E_{\rm ECL}$) in the electromagnetic calorimeter, computed in the CM frame and not associated with neutral particles or charged tracks used in the 
$B_{\rm tag}$ reconstruction,
is used to discriminate between signal and background.
For $B^0\to invisible+\gamma$, the energy of the highest-energy photon remaining in the event 
is also removed from the $E_{\rm extra}$ computation. In signal events $E_{\rm extra}$ is
strongly peaked at zero, while for the background its distribution increases 
uniformly with $E_{\rm extra}$.
The main difference between the Belle and \babar\ analyses is that Belle performs a full 
reconstruction of the $B_{\rm tag}$ candidate in hadronic final states, while \babar\ opts for a partial 
reconstruction in the $D^{(*)-}\ell^+\nu$ decays. The latter has the advantage of a higher 
reconstruction efficiency with respect to the former (by a factor $\approx 10$: the $B\to invisible$ 
efficiency is 0.18\% for \babar\ and 0.022\% for Belle), but has the disadvantage of the 
presence of the invisible neutrino, which prevents the exploitation of kinematic variables such as the 
reconstructed $B^0$ mass. $S/B$ is therefore worse in the case of the partial $B_{\rm tag}$
reconstruction, though the background contamination is mitigated by the presence of a high 
momentum lepton.

\begin{figure*}[!htbp]
\centering
\includegraphics[width=0.48\textwidth]{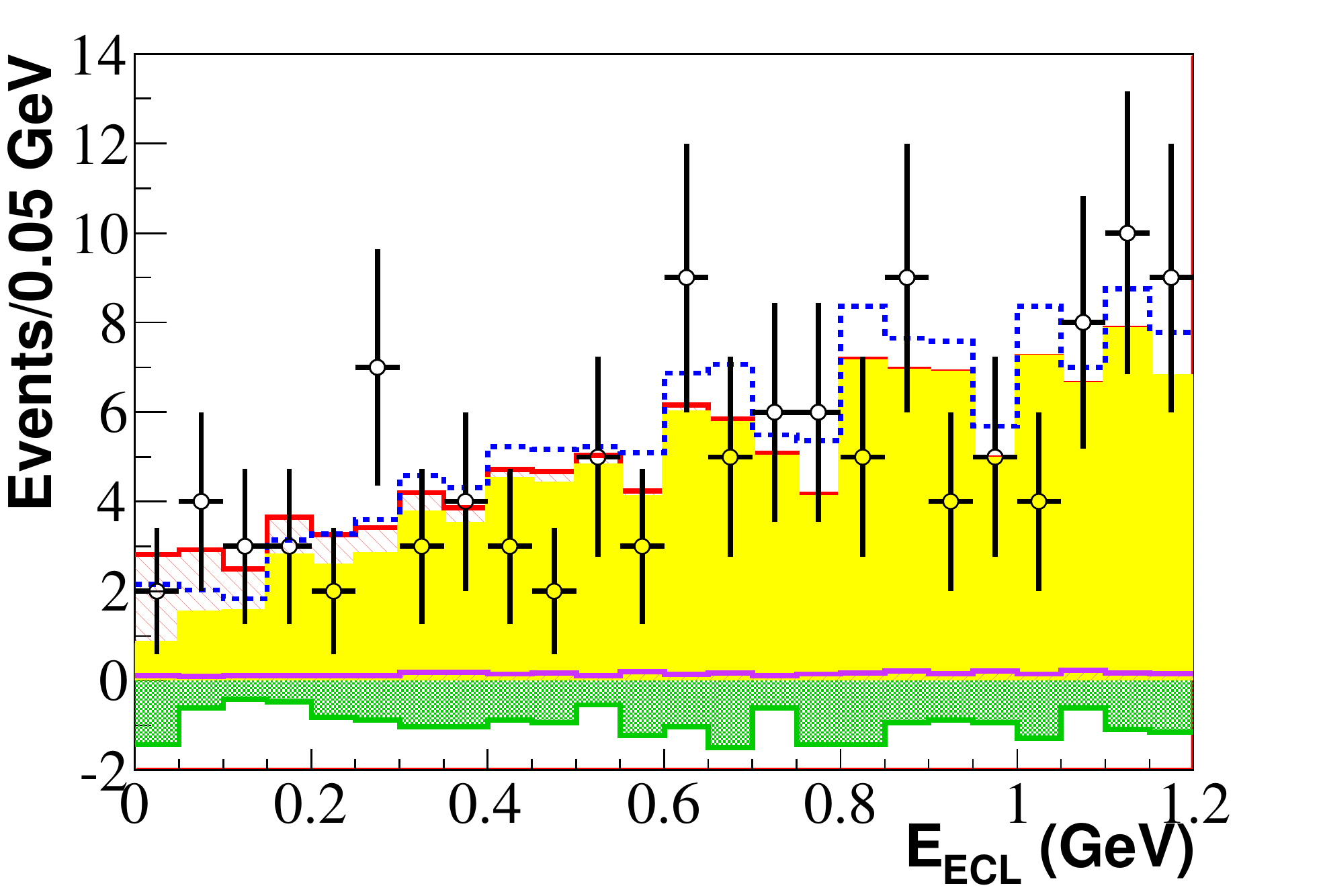}
\caption{$E_{\rm ECL}$ distribution of $B \to invisible$ candidates (points) selected by Belle, with fit results superimposed. The red cross-hatched region is the signal component, on the top of the total background shown in the yellow filled histogram. The blue dashed curve is the generic B contribution, which is larger than the total because of the negative fit result for the non-B background shown in the green dotted histogram. The purple hatched area corresponds to the rare $B$ contribution.} 
\label{fig:Belle:invisible}
\end{figure*}

\begin{figure*}[!htbp]
\centering
\includegraphics[width=0.35\textwidth,height=5cm]{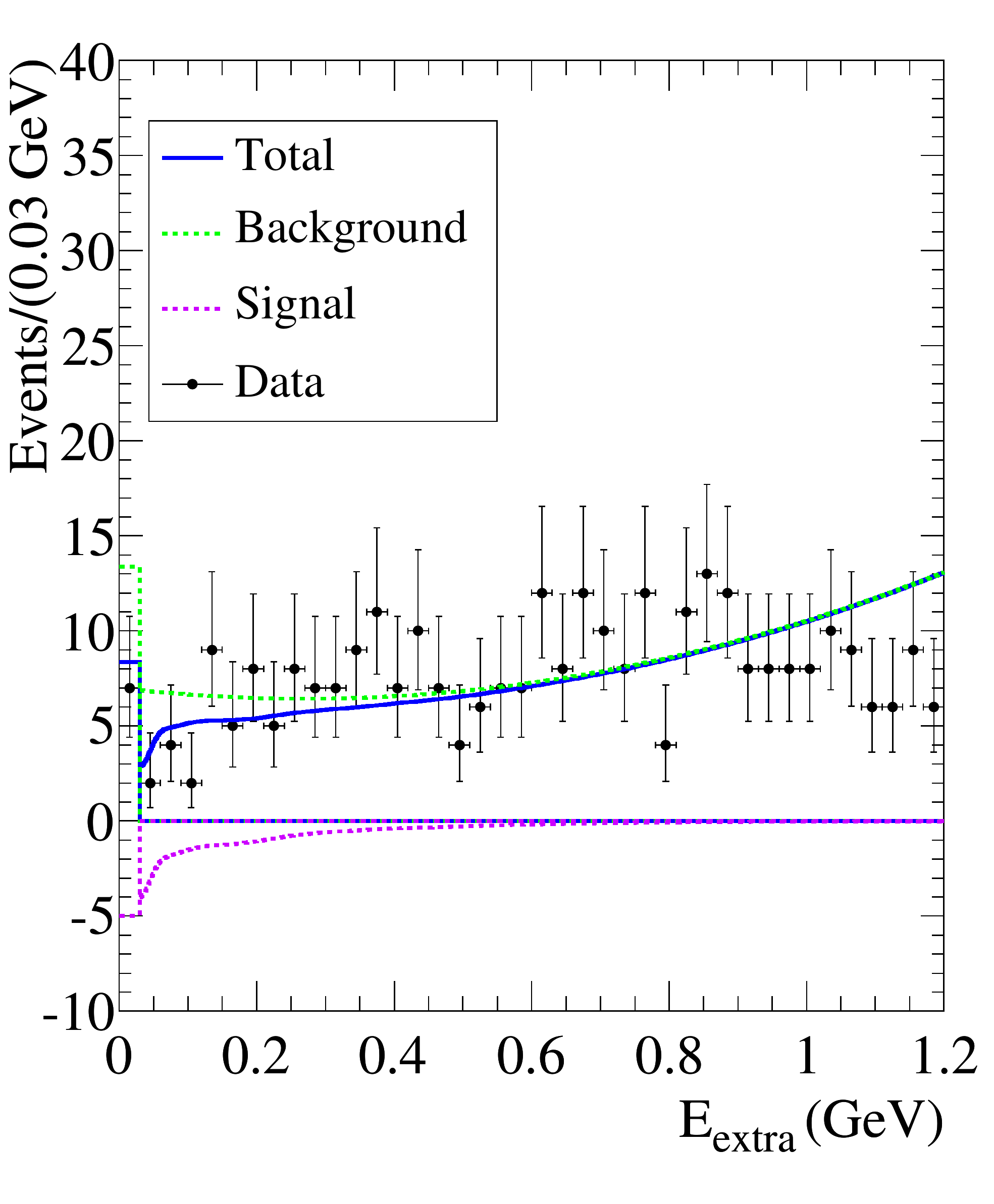}
\hspace{1cm}
\includegraphics[width=0.35\textwidth,height=5cm]{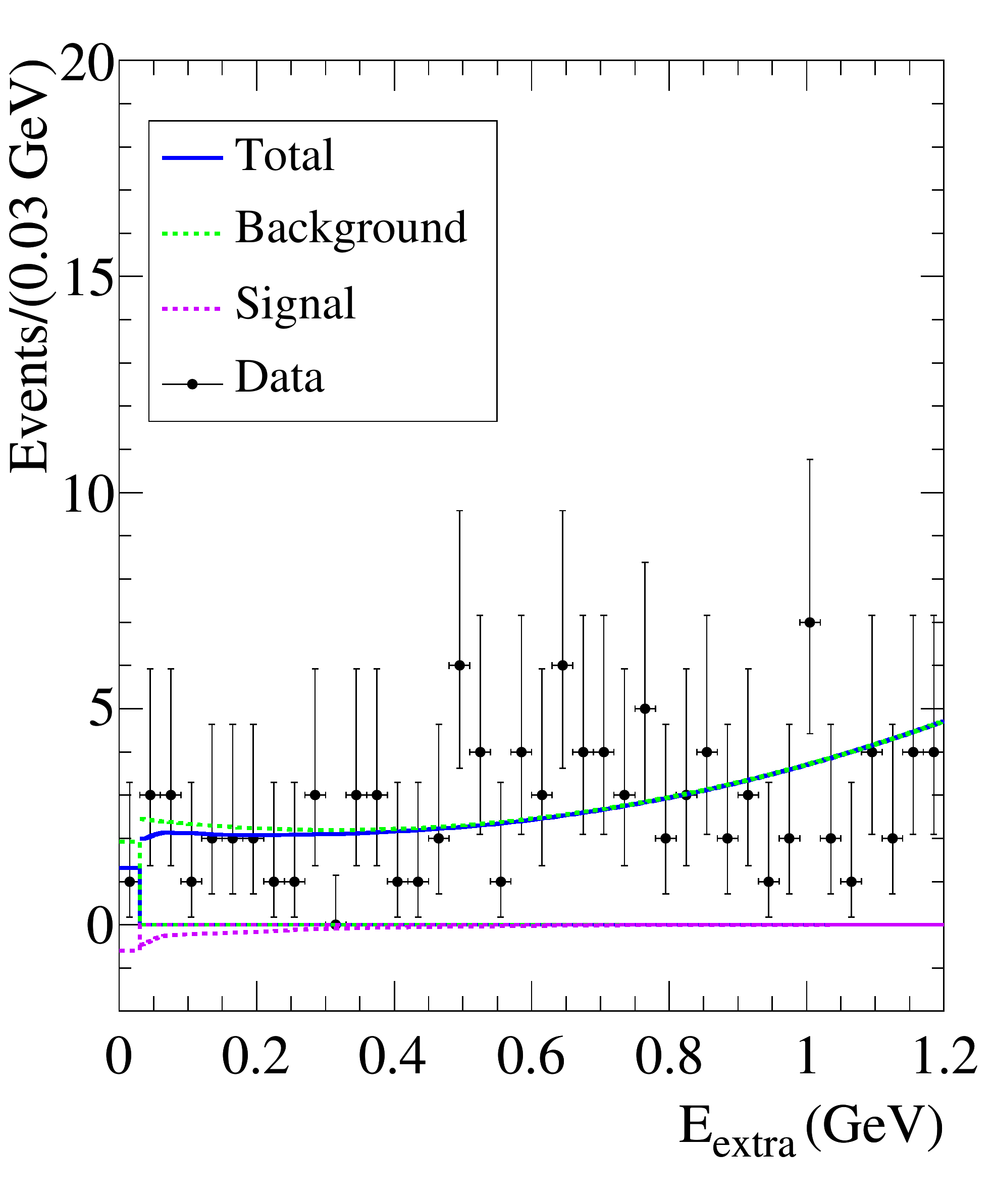}
\caption{$E_{\rm extra}$ distributions of $B \to invisible$ (left) and $B\to invisible + \gamma$ (right) candidates selected by \babar\ (points), with fit results superimposed: signal (purple dotted line),
background (green, dotted line) and sum of signal and background (blue solid line).}
\label{fig:BaBar:invisible}
\end{figure*}

The signal yield is extracted from a maximum likelihood fit to the
two-dimensional distributions of $E_{\rm ECL}$ and of an event-shape
variable, $\cos\theta_B$ (Belle), or to the one-dimensional $E_{\rm extra}$ distribution (\babar),
 as illustrated in Figures~\ref{fig:Belle:invisible} and~\ref{fig:BaBar:invisible}.
No evidence for $B\to invisible (+\gamma)$ decays is found, and 90\% confidence level upper limits 
are set -- with a Bayesian technique -- on each branching fraction,
at the level of a few times $10^{-5}$, as summarized in Table~\ref{tab:invisible}. The Belle limit is 
a factor $\approx 5$ worse than \babar\ because of the lower efficiency of the hadronic tag
full reconstruction. 

\begin{table}[!htbp]
\begin{center}
\caption{90\% C.L. upper limits on the branching fractions of $B\to invisible (+ \gamma)$ decays 
measured by the Belle and \babar\ collaborations~\cite{bib:Belle:invisible,bib:BaBar:invisible}.}
\begin{tabular}{|l|c|c|}
\hline \textbf{Decay} & \textbf{BF (Belle)} & \textbf{BF (\sbabar)} \\
\hline $B^0\to invisible$ & $<13 \times 10^{-5}$ & $<2.4 \times 10^{-5}$ \\
\hline $B^0\to invisible + \gamma$ & $-$ & $<1.7\times 10^{-5}$ \\
\hline
\end{tabular}
\label{tab:invisible}
\end{center}
\end{table}

\section{CONCLUSION}
Several results on rare $B$ decays obtained at the $B$ factories have been
discussed. The main results are:
\begin{itemize}
\item \CP violation has been observed with $>3\sigma$ significance in $B^\pm\to \eta K^\pm$
and $B^\pm \to \eta \pi^\pm$. Large \CP asymmetries are observed. $B^0{\to}\eta K^0$ is 
observed for the first time, with a branching fraction smaller than that of $B^+{\to}\eta K^+$.
\item \CP-violation parameters in $B^0\to K^0_SK^0_SK^0_S$ are
compatible within two standard deviations with those measured in
tree-dominated modes such as $B^0 \to J/\psi K_S^0$, as expected in the SM. 
\CP conservation is excluded for the first time in these decay with a significance of more than $3\sigma$, including systematic uncertainties.
\item The \CP asymmetry in $B^+\to\phi K^+$ differs from 0 by $2.8\sigma$, and is in 
slight tension with SM predictions ($0-4.7\%$). 
\item The amount of \CP violation in $B^0\to \phi K^0_S$ and in other
$B^0\to K^+K^-K^0_S$ decays is in excellent agreement with SM expectations: $\beta_{\rm eff}$ is 
very close to $\beta$ measured in $B$ decays to charmonium, and $\mathcal{C}$ is consistent with
zero.
\item No evidence of \CP violation has been found in $B^\pm \to K^*(892)^\pm\pi^0$, though
the large uncertainty on the \CP asymmetry does not exclude \CP-violating effects as large as 
$20-30\%$, as expected in the SM.
\item No evidence of lepton-flavor violating $B^\pm \to h^\pm \tau \ell$ decays has been found,
and upper limits of a few times $10^{-5}$ on their branching fractions have been set, improving
by a factor 5 previous limits on $\mu-\tau$ lepton-flavor violating effective couplings.
\item No evidence of $B\to invisible (+\gamma)$ decays has been found, and upper limits of a 
few times $10^{-5}$ on their branching fractions have been set, 1--2 order of magnitude 
still higher than the values predicted by some theoretical models beyond the SM.
\end{itemize}
All the results are in agreement with Standard Model expectations: no evidence of new physics
is found.


\begin{thebibliography}{99} % Use for 10-99 references

\bibitem{bib:Cabibbo}
N. Cabibbo, ``Unitary Symmetry and Leptonic Decays", Phys. Rev. Lett. {\bf 10}, 531 (1963).

\bibitem{bib:KM}
M. Kobayashi, T. Maskawa, ``\CP violation in the renormalizable theory of weak interaction",
Prog. Theor. Phys. {\bf 49}, 652 (1973).

\bibitem{bib:HFAG}
Heavy Flavor Averaging Group,
    {\tt http://www.slac.stanford.edu/xorg/hfag/}

\bibitem{bib:theo:etah1}
H. J. Lipkin, 
``Interference effects in $K\eta$ and $K\eta^\prime$ decay modes of heavy mesons. Clues to understanding weak transitions and \CP violation",
Phys. Lett. B {\bf 254}, 247 (1991).

\bibitem{bib:theo:etah2}
M. Bander, D. Silverman, and A. Soni, 
``\CP Noninvariance in the Decays of Heavy Charged Quark Systems",
Phys. Rev. Lett. {\bf 43}, 242 (1979).

\bibitem{bib:theo:etah3}
H.-Y. Cheng and C.-K. Chua, 
``Revisiting charmless hadronic $B_{u,d}$ decays in QCD factorization",
Phys. Rev. D {\bf 80}, 114008 (2009).

\bibitem{bib:theo:etah4}
  Z.~J.~Xiao, Z.~Q.~Zhang, X.~Liu, and L.~B.~Guo,
  ``Branching ratios and \CP asymmetries of $B \to K \eta^{(\prime)}$ decays in
  the pQCD approach,'' Phys.\ Rev.\  D {\bf 78}, 114001 (2008).

\bibitem{bib:theo:etah5}
  H.~S.~Wang, X.~Liu, Z.~J.~Xiao, L.~B.~Guo, and C.~D.~Lu,
  ``Branching ratio and \CP asymmetry of $B \to \pi \eta^{(\prime)}$ decays in the
  perturbative QCD approach,''
  Nucl.\ Phys.\  B {\bf 738}, 243 (2006).

\bibitem{bib:theo:etah6}
A.G. Akeroyd, C.H. Chen, and C.Q. Geng,
``$B\to\eta^{(\prime)}(l^-\bar\nu,l^+l^-,K,K^*)$ decays in the quark-flavor mixing scheme",
 Phys. Rev. D {\bf 75}, 054003 (2007).

\bibitem{bib:theo:etah7}
  A.~R.~Williamson and J.~Zupan,
  ``Two body $B$ decays with isosinglet final states in SCET,''
  Phys.\ Rev.\  D {\bf 74}, 014003 (2006)
  [Erratum-ibid.\  D {\bf 74}, 039901 (2006)].

\bibitem{bib:theo:etah8}
M.~Beneke and M.~Neubert,
``Flavor-singlet $B$ decay amplitudes in QCD factorization,''
Nucl.\ Phys.\  B {\bf 651}, 225 (2003).

\bibitem{bib:theo:etah9} 
  C.-W.~Chiang, M.~Gronau, J.~L.~Rosner and D.~A.~Suprun,
  ``Charmless $B \to P P$ decays using flavor SU(3) symmetry,''
  Phys.\ Rev.\ D {\bf 70}, 034020 (2004).

\bibitem{bib:Belle:etah}
Belle Collaboration, C.-T. Hoi {\em et al.}, 
``Evidence for direct \CP violation in $B^\pm \to \eta h^\pm$ and observation of $B^0\to \eta K^0$", 
Phys. Rev. Lett. {\bf 108}, 031801 (2012).

\bibitem{bib:sin2betaeff1}
H.-Y. Cheng, C.-K. Chua, and A. Soni
``$CP$-violating asymmetries in ${B}^{0}$ decays to ${K}^{+}{K}^{-}{K}_{S(L)}^{0}$ and ${K}_{S}^{0}{K}_{S}^{0}{K}_{S(L)}^{0}$", Phys. Rev. D {\bf 72}, 094003 (2005).

\bibitem{bib:sin2betaeff2}
M. Beneke, ``Corrections to $\sin 2\beta$ from \CP asymmetries in $B^0\to(\pi^0,\rho^0,\eta,\eta^\prime,\omega,\phi)K^0_S$ decays", Phys. Lett. B {\bf 620}, 143 (2005).

\bibitem{bib:sin2betaeff3}
H.-n. Li and S. Mishima, ``Penguin-dominated $B\to PV$ decays in NLO perturbative QCD",
Phys. Rev. D {\bf 74}, 094020 (2006).

\bibitem{bib:BaBar:KsKsKs} 
\babar\ Collaboration, J.~P.~Lees {\em et al.}, 
  ``Amplitude analysis and measurement of the time-dependent CP asymmetry of $B^0 \to K_S^0 K_S^0 K_S^0$ decays,'' 
  Phys.\ Rev.\ D {\bf 85}, 054023 (2012).
  
\bibitem{bib:fX}
\babar\ Collaboration, B.~Aubert {\it et al.},
  ``Dalitz plot analysis of the decay $B^\pm \to K^\pm K^\pm K^\mp$,''
  Phys.\ Rev.\ D {\bf 74}, 032003 (2006).

\bibitem{bib:BaBar:KKK} 
\babar\ Collaboration, J.~P.~Lees {\em et al.}, 
 ``Study of \CP violation in Dalitz-plot analyses of $B^0 \to K^+K^-K^0_S$, $B^+ \to K^+K^-K^+$, and 
 $B^+ \to K^0_SK^0_SK^+$,''  Phys.\ Rev.\ D {\bf 85}, 112010 (2012).
  
\bibitem{bib:theo:Kpi0pi01}
 Q. Chang, X.-Q. Li, and Y.-D. Yang, ``Revisiting $B \to \pi K$, $\pi K^*$ and $\rho K$ decays: \CP violations and implication for new physics",
 J. High Energy Phys. {\bf 09} (2008) 038.

\bibitem{bib:theo:Kpi0pi02}
C.-W. Chiang and D. London, ``Looking for new physics in $B \to K^*\pi$ and $B\to \rho K$ decays", Mod. Phys. Lett. A {\bf 24}, 1983 (2009).

\bibitem{bib:theo:Kpi0pi03}
M. Gronau, D. Pirjol, and J. Zupan, ``\CP asymmetries in $B{\to}K\pi, K^*\pi, \rho K$ decays", Phys. Rev. D {\bf 81}, 094011 (2010).
 
\bibitem{bib:BaBar:Kpi0pi0}
\babar\ Collaboration, J.~P.~Lees {\em et al.}, ``Observation of the rare decay $B^+\to K^+\pi^0\pi^0$ 
and measurement of the quasi-two-body contributions $B^+\to K^*(892)^+\pi^0$, 
$B^+\to f_0(980)K^+$, and $B^+\to \chi_{c0}K^+$", Phys. Rev. D {\bf 84}, 092007 (2011).

\bibitem{bib:theory:htaul1}
X.-G. He, G. Valencia, and Y. Wang, ``Lepton flavor violating $\tau$ and $B$ decays and heavy neutrinos", Phys. Rev. D {\bf 70}, 113011 (2004).

\bibitem{bib:theory:htaul2}
M. Sher and Y. Yuan, ``Rare $B$ decays, rare $\tau$ decays, and grand unification", Phys. Rev. D {\bf 44}, 1461 (1991).

\bibitem{bib:theory:htaul3} 
D. Black, T. Han, H.-J. He, and M. Sher, ``$\tau-\mu$ flavor violation as a probe of the scale of new physics", Phys. Rev. D {\bf 66}, 053002 (2002).

\bibitem{bib:theory:htaul4}
T. Fujihara, S.K. Kang, C. S. Kim, D. Kimura, and T. Morozumi,
``Low scale seesaw model and lepton flavor violating rare $B$ decays", 
Phys. Rev. D {\bf 73}, 074011 (2006).

\bibitem{bib:BaBar:htaul}
\babar\ Collaboration, J.~P.~Lees {\em et al.}, ``Search for the decay modes $B^\pm \to h^\pm \tau \ell$," Phys. Rev. D {\bf 86}, 012004 (2012).

\bibitem{bib:theo:invisible1}
G. Buchalla and A. J. Buras,
``QCD corrections to rare $K$ and $B$ decays for arbitrary top quark mass",
Nucl. Phys. B {\bf 400}, 225 (1993).

\bibitem{bib:theo:invisible2}
C. D. Lu and D. X. Zhang, 
``$B_s (B_d) \to \gamma \gamma \bar \nu$",
  Phys.\ Lett.\ B {\bf 381}, 348 (1996).

\bibitem{bib:theo:invisible3}
A. Dedes, H. Dreiner, and P. Richardson, 
``Attempts at explaining the NuTeV observation of dimuon events",
Phys. Rev. D {\bf 65}, 015001 (2002).

\bibitem{bib:theo:invisible4}
K. Agashe and G.-H. Wu, 
``Remarks on models with singlet neutrino in large extra dimensions",
Phys. Lett. B {\bf 498}, 230 (2001).

\bibitem{bib:theo:invisible5}
H. Davoudiasl, P. Langacker, and M. Perelstein, 
``Constraints on large extra dimensions from neutrino oscillation experiments",
Phys. Rev. D {\bf 65}, 105015 (2002).

\bibitem{bib:Belle:invisible}
Belle Collaboration, C.~L.~Hsu {\it et al.}, ``Search for $B^{0}$ decays to invisible final states,''
  Phys.\ Rev.\ D {\bf 86}, 032002 (2012).

\bibitem{bib:BaBar:invisible} 
\babar\ Collaboration, J.~P.~Lees {\it et al.},
  ``Improved Limits on $B^0$ Decays to Invisible Final States and to $\nu \bar{\nu} \gamma$,''
  arXiv:1206.2543 [hep-ex]. Submitted to Phys. Rev. D.

\end{thebibliography}
\end{document}